\pdfoutput=1
\documentclass{article}

\usepackage[utf8]{inputenc}
\usepackage{bm}
\usepackage{graphicx}
\usepackage{algorithm}
\usepackage[noend]{algpseudocode}
\usepackage{geometry}
\usepackage{graphicx,tikz}
\usepackage{subfig}
\usepackage{hyperref}
\usetikzlibrary{shapes}
\usepackage{float}

\begin{document}

\title{A Scalable, Trustworthy Infrastructure for Collaborative Container Repositories}
\author{Franklin Wei, Mahalingam Ramkumar, Stephen R. Tate and Somya D Mohanty} 
\date{}
\maketitle





\date{May 2018}

\begin{abstract}

We present a scalable ``Trustworthy Container Repository'' (TCR) infrastructure for the storage of software container images, such as those used by Docker. Using an authenticated data structure based on index-ordered Merkle trees (IOMTs), TCR aims to provide assurances of 1) Integrity, 2) Availability, and 3) Confidentiality to its users, whose containers are stored in an untrusted environment. Trust within the TCR architecture is rooted in a low-complexity, tamper-resistant trusted module. The use of IOMTs allows such a module to efficiently track a virtually unlimited number of container images, and thus provide the desired assurances for the system's users. Using a simulated version of the proposed system, we demonstrate the scalability of platform by showing logarithmic time complexity up to $2^{25}$ (32 million) container images. This paper presents both algorithmic and proof-of-concept software implementations of the proposed TCR infrastructure.


\end{abstract}

\maketitle

\section{Introduction}

Recent years have seen the rise of ``containerization'' \cite{vaughan2006new} software such as Docker, which facilitates the modular development and deployment of software applications. Such software often depends on a centralized repository (e.g. Docker Hub), for storing and distributing container images. Because containers contain code that is executed on client machines, these centralized repositories present an appealing attack vector to potential bad actors. Malicious entities can use the implicit trust placed in the hardware, software, and even the administrative personnel of such repository services as an starting point for conducting attacks against the users of the service. As a result, there is a need for a trustworthy architecture capable of provisioning explicit trust in the operations of the repository.

In a traditional repository service, users communicate with a untrusted repository service \textbf{\textit{S}}, which provides the basic services of a repository server: creating containers, modifying containers, and retrieving contents. Under such a model, users have no way of verifying that \textit{\textbf{S}} behaves properly; i.e. \textit{\textbf{S}} could tamper with the data entrusted to it, and/or improperly deny service by falsely claiming that requested containers do not exist, and users would have no way to learn of the misbehavior.

In order to address such limitations in ensuring trust in traditional models, we present a Trustworthy Container Repository (TCR) infrastructure. TCR bootstraps security assurances from a low-complexity trusted module \textbf{T}, and amplifies its trustworthiness using an authenticated data structure towards the operation of the untrusted repository service \textbf{\textit{S}}. The TCR model uses a variant of the classic Merkle tree, an index-ordered Merkle tree (IOMT), as an authenticated data structure to efficiently track a large number of container records. Based on the model, assurances of container integrity, availability, and confidentiality are provided to users of \textbf{\textit{S}}.

TCR differs from traditional models by introducing the trusted module \textbf{T}, which acts as a ``gatekeeper.'' The presence of \textbf{T} ensures that all operations on the container repository are properly authenticated, and that misbehavior by \textit{\textbf{S}} is immediately obvious to users. Users do not communicate directly with the trusted module \textbf{T}. Instead, \textit{\textbf{S}} is expected to act as an intermediary between users and \textbf{T} to provide validity of its operations. All user requests are relayed by \textbf{\textit{S}} to \textbf{T}, which uses simple cryptographic methods and self-memoranda (\emph{certificates}) to perform operations on the IOMT data structure. These methods allow \textbf{T} to track the state of the container repository, given by the root of the IOMT stored within its trusted, tamper-resistant boundary. This system ensures detection of any illegal operations on the state of the repository, even though almost all data is stored by the untrusted service \textit{\textbf{S}}.

Responses to requests from users are given by \textit{\textbf{S}} and are verifiable with a \emph{proof of trust} from \textbf{T} (users and \textbf{T} share keys for verification). For example, for query requests (content and information retrieval), \textbf{T} certifies that the information returned by \textit{\textbf{S}} is up-to-date and reflects the true state of the container repository. \textbf{T} also certifies that update requests (container creation and modification), which modify the state of the repository, are reflected in the internal state of the module. The TCR model ensures \textbf{T} will refuse to issue proofs of trust for information inconsistent with the latest repository state.

As a part of this paper, we also develop and evaluate a proof-of-concept implementation for the TCR infrastructure. The implementation is based on a client-server architecture, and simulates the operation of \textbf{T} and \textbf{\textit{S}}. A SQL-based database is used to maintain the container records and IOMT data structure. The evaluation explores the performance scalability of the model in large container repositories (1024 - 32 million containers).

The paper is organized as follows. We begin by discussing containerization and Docker in Section \ref{docker}, and index-ordered Merkle trees (IOMTs), a key component of TCR, in Section \ref{iomt}. We review related work in the field and outline the current security issues of Docker Hub in Section \ref{related}. In Section \ref{tcroverview}, we present a high-level overview of TCR, describing the interaction of the various entities involved. Descriptions of its various components and the underlying algorithms are given in Sections \ref{certificates}--\ref{protocols}. In Section \ref{implementation}, we present our preliminary implementation of TCR. Section \ref{results} discusses the observed performance results (Section \ref{perfeval}) and evaluates the security analysis (\ref{seceval}) of the implemented model. Finally, in Section \ref{conclusion}, we summarize our contributions and lay out possibilities for future work.

\section{Background}
\label{background}

\subsection{Containerization and Docker}
\label{docker}

Containerization \cite{vaughan2006new, dua2014virtualization} is the use of lightweight virtualization architectures for software packaging and deployment. Similar to a traditional virtual machine (VM), containerization combines application code (say, for a web server), libraries, and configuration files into an object called an \emph{image}. Images can then used to instantiate containers, which are virtualized environments in which applications can run.

Whereas a traditional VM must virtualize the entire software stack from the operating system up, containers are extremely lightweight because they provide a higher level of virtualization: while VMs provide low-level, instruction-scale virtualization, containers run with only a thin layer of abstraction separating them from the host operating system, reducing execution overhead. Additionally, container images are smaller than similar VM images, making them more efficient, space-wise, for moving applications between cloud providers.

Docker \cite{merkel2014docker} is currently one of the most widely used models of containerization. Docker containers use operating system features such as Linux Containers (LXC) \cite{linux_containers} and Control Groups (cgroups) \cite{c_groups} to provide the necessary isolation between containers and the host machine.

Within the Docker architecture, a file called a $Dockerfile$ is used to build a Docker container image. More specifically, the $Dockerfile$ is a build file containing all instruction/commands to be executed in sequence in order to create a new container $image$. The file contains all the necessary code, library, data, and initialization scripts to enable the container operation. Deployment of the container $image$ for operation is done using a $Composefile$ (or stack file), which contains instructions (written in YAML) for configuration of container application services. The $Dockerfile$, $Composefile$, and the corresponding container $image$ form the key components of a container based repository in our approach.

\subsection{Merkle Tree}

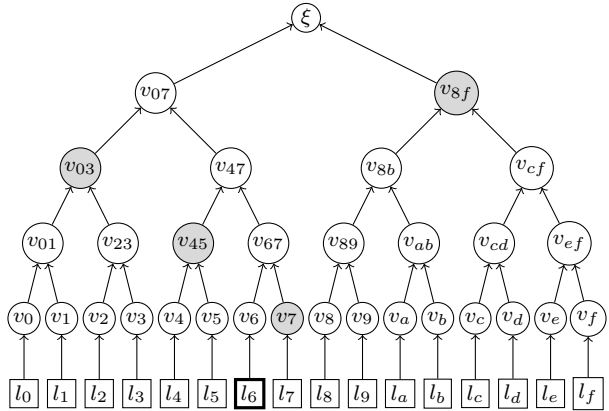
\begin{figure}[!h]
\begin{center}
 \begin{tikzpicture}[scale=1]

\tikzstyle{every node}=[draw,rectangle,inner sep=2pt]
\draw (0,0) node (l0) {\footnotesize $ l_0$};
\draw (0.5,0) node (l1) {\footnotesize $l_1$};
\draw (1,0) node (l2) {\footnotesize $l_2$};
\draw (1.5,0) node (l3) {\footnotesize $l_3$};
\draw (2,0) node (l4) {\footnotesize $l_4$};
\draw (2.5,0) node (l5) {\footnotesize $l_5$};
\draw[very thick] (3,0) node (l6) {\footnotesize $l_6$};
\draw (3.5,0) node (l7) {\footnotesize $l_7$};
\draw (4,0) node (l8) {\footnotesize $l_8$};
\draw (4.5,0) node (l9) {\footnotesize $l_9$};
\draw (5,0) node (la) {\footnotesize $l_a$};
\draw (5.5,0) node (lb) {\footnotesize $l_b$};
\draw (6,0) node (lc) {\footnotesize $l_c$};
\draw (6.5,0) node (ld) {\footnotesize $l_d$};
\draw (7,0) node (le) {\footnotesize $l_e$};
\draw (7.5,0) node (lf) {\footnotesize $l_f$};

\tikzstyle{every node}=[draw,circle,inner sep=1pt]

\draw (0,1) node (v0) {\footnotesize $v_0$};
\draw (0.5,1) node (v1) {\footnotesize $v_1$};
\draw (1,1) node (v2) {\footnotesize $v_2$};
\draw (1.5,1) node (v3) {\footnotesize $v_3$};
\draw (2,1) node (v4) {\footnotesize $v_4$};
\draw (2.5,1) node (v5) {\footnotesize $v_5$};
\draw (3,1) node (v6) {\footnotesize $v_6$};
\draw[color=black!100] (3.5,1) node [fill=gray!30] (v7) {\footnotesize $v_7$};
\draw (4,1) node (v8) {\footnotesize $v_8$};
\draw (4.5,1) node (v9) {\footnotesize $v_9$};
\draw (5,1) node (va) {\footnotesize $v_a$};
\draw (5.5,1) node (vb) {\footnotesize $v_b$};
\draw (6,1) node (vc) {\footnotesize $v_c$};
\draw (6.5,1) node (vd) {\footnotesize $v_d$};
\draw (7,1) node (ve) {\footnotesize $v_e$};
\draw (7.5,1) node (vf) {\footnotesize $v_f$};

\draw[->]  (l0) -- (v0);
\draw[->]  (l1) -- (v1);
\draw[->]  (l2) -- (v2);
\draw[->]  (l3) -- (v3);
\draw[->]  (l4) -- (v4);
\draw[->]  (l5) -- (v5);
\draw[->]  (l6) -- (v6);
\draw[->]  (l7) -- (v7);
\draw[->]  (l8) -- (v8);
\draw[->]  (l9) -- (v9);
\draw[->]  (la) -- (va);
\draw[->]  (lb) -- (vb);
\draw[->]  (lc) -- (vc);
\draw[->]  (ld) -- (vd);
\draw[->]  (le) -- (ve);
\draw[->]  (lf) -- (vf);

\draw (0.25,2) node (v01) {\footnotesize $v_{01}$};
\draw (1.25,2) node (v23) {\footnotesize $v_{23}$};
\draw[color=black!100] (2.25,2) node [fill=gray!30] (v45) {\footnotesize $v_{45}$};
\draw (3.25,2) node (v67) {\footnotesize $v_{67}$};
\draw (4.25,2) node (v89) {\footnotesize $v_{89}$};
\draw (5.25,2) node (vab) {\footnotesize $v_{ab}$};
\draw (6.25,2) node (vcd) {\footnotesize $v_{cd}$};
\draw (7.25,2) node (vef) {\footnotesize $v_{ef}$};

\draw[->]  (v0) -- (v01);
\draw[->]  (v1) -- (v01);
\draw[->]  (v2) -- (v23);
\draw[->]  (v3) -- (v23);
\draw[->]  (v4) -- (v45);
\draw[->]  (v5) -- (v45);
\draw[->]  (v6) -- (v67);
\draw[->]  (v7) -- (v67);
\draw[->]  (v8) -- (v89);
\draw[->]  (v9) -- (v89);
\draw[->]  (va) -- (vab);
\draw[->]  (vb) -- (vab);
\draw[->]  (vc) -- (vcd);
\draw[->]  (vd) -- (vcd);
\draw[->]  (ve) -- (vef);
\draw[->]  (vf) -- (vef);

\draw[color=black!100] (0.75,3) node [fill=gray!30] (v03) {\footnotesize $v_{03}$};
\draw (2.75,3) node (v47) {\footnotesize $v_{47}$};
\draw (4.75,3) node (v8b) {\footnotesize $v_{8b}$};
\draw (6.75,3) node (vcf) {\footnotesize $v_{cf}$};

\draw[->]  (v01) -- (v03);
\draw[->]  (v23) -- (v03);
\draw[->]  (v45) -- (v47);
\draw[->]  (v67) -- (v47);
\draw[->]  (v89) -- (v8b);
\draw[->]  (vab) -- (v8b);
\draw[->]  (vcd) -- (vcf);
\draw[->]  (vef) -- (vcf);

\draw (1.75,4) node (v07) {\footnotesize $v_{07}$};
\draw[color=black!100] (5.75,4) node [fill=gray!30] (v8f) {\footnotesize $v_{8f}$};

\draw[->]  (v03) -- (v07);
\draw[->]  (v47) -- (v07);
\draw[->]  (v8b) -- (v8f);
\draw[->]  (vcf) -- (v8f);

\draw (3.75,5) node (v0f) {\footnotesize $\xi$};

\draw[->]  (v07) -- (v0f);
\draw[->]  (v8f) -- (v0f);

\end{tikzpicture}
\end{center}
\caption{A binary Merkle tree with 16 leaves ($h=4$). The complementary nodes of the leaf record $l_6$ are $v_7$, $v_{45}$, $v_{03}$ and $v_{8f}$ (shaded).}
\label{fig:mtree}
\end{figure} 

A Merkle tree \cite{merkle88}, also called a ``hash tree'', is a tree data structure whose internal nodes are the cryptographic hash ($h()$, where $h$ can be SHA-1, SHA-2, etc.) of its child nodes. A common variant of the Merkle tree is the binary tree, in which each internal node has a maximum of two child nodes. In such a tree of height $h$, the data structure consists of $N = 2^h$ leaf nodes at the lowest level. Figure \ref{fig:mtree} shows an example Merkle tree of height $h=4$ with $2^4 = 16$ leaf nodes.

Levels of a tree are numbered with the lowest level (with the most nodes) as $L = 0$, and the root level (with only one node) as $L = h$. We draw an important distinction between \emph{leaf nodes} and \emph{leaf records}: a node of the tree at the level $L = 0$ is termed a \emph{leaf node}, and has the value $v_n = h(l_n)$, where $l_n$ is the value of the node's corresponding \emph{leaf record}. In an regular Merkle tree, the form of leaf records is unconstrained; however, in an index-ordered Merkle tree (IOMT), described below, they are constrained to a fixed form.

In order to calculate the value of the parent nodes, a function $F_{parent}()$ (Algorithm \ref{alg:merkle_par}) takes the values of the two child nodes ($v_i$ and $v_j$) and their orientation in the tree (e.g. $order_i = LEFT$, meaning first value $v_i$ is the left child) as its parameters.

If both $v_i$ and $v_j$ are nonzero, the value of the parent is then given by the hash ($h()$) of the concatenated child node values. However, if one or more of the child nodes has a zero value, the parent retains the value of the other node (which might also be zero). (Giving the hash value of all zeros a special meaning is safe because it is computationally infeasible to find a preimage $v$ so that $h(v) = 0$ with a well-designed $h()$.)

In other words, each node at level $L$ (where $0 \leq L \leq h$) is mapped to its parent at level $L+1$, ending in the root of the tree $\xi$. The root node can be viewed as a single, compact cryptographic commitment to all nodes and leaf records of the tree.

\begin{algorithm}
\caption{Merkle tree parent calculation}\label{f_parent}
\label{alg:merkle_par}
\begin{algorithmic}[1]
\Procedure{$F_{parent}$}{$v_i, v_j, order_i$}

\If{$v_i = 0$} \State \Return{$v_j$} \Comment {$v_j$ can be zero as well}
\ElsIf{$v_j = 0$} \State \Return{$v_i$}
\Else \Comment {Both $v_i$ and $v_j$ nonzero}
\If{$order_i = LEFT$}
\State \Return {$h(v_i \parallel v_j)$}
\ElsIf{$order_i = RIGHT$}
\State \Return {$h(v_j \parallel v_i)$}
\EndIf
\EndIf

\EndProcedure
\end{algorithmic}
\end{algorithm}

One of the key properties of a Merkle tree is that every individual leaf record can be verified (proving that it exists in a tree with a given root) or updated with $h + 1$ operations performed on the tree. Since $N = 2^h$, these operations take $O(log N)$ time.

More specifically, for a verification of a leaf record, a set of complementary nodes from the tree can be provided to map its value to the root $\xi$. In our example tree (Figure \ref{fig:mtree}), for verification of the leaf record $l_6$, the set of complementary nodes is

\[
[X_{comp}] = [(v_7, RIGHT), (v_{45}, LEFT), (v_{03}, LEFT), (v_{8f}, RIGHT)].
\]

The root of the tree can then be calculated with the operations, $v_6 = h(l_6)$; $v_{67} = h(v_6 \parallel v_7); v_{47} = h(v_{45} \parallel v_{67}); v_{07} = h(v_{03} \parallel v_{47})$; and finally $\xi = h(v_{07} \parallel v_{8f})$.

Function $F_{mt}()$ (Algorithm \ref{alg:merkle_root}) describes the general method for calculating the root of a Merkle tree. The input parameters given to it are the leaf node $X$, the list of its complementary nodes $[X_{comp}]$, and a list $[X_{orders}]$ indicating the ordering of the nodes.

\begin{algorithm}
\caption{Merkle tree root calculation procedure}\label{f_bt}
\label{alg:merkle_root}
\begin{algorithmic}[1]
\Procedure{$F_{mt}$}{$X, [X_{comp}], [X_{orders}]$}
\State{$Y \gets X$}

\For{$I$ \textbf{in} $[1 .. X_{comp}]$}
\State{$Y \gets F_{parent}([X_{comp}]_I, Y, [X_{orders}]_I)$}
\EndFor

\State \Return{Y}

\EndProcedure
\end{algorithmic}
\end{algorithm}

\subsection{Index-Ordered Merkle Tree}

\label{iomt}

Although a ordinary Merkle tree enables trust in the values of the leaves with a single root $\xi$, where $\xi$ can be stored in a secure boundary to mitigate manipulation, it is unable to prevent malicious replay attacks; a malicious entity could keep duplicate instances of leaves and replay incorrect information based on older leaf values. An ordinary Merkle tree is also limited in its ability to prove the \textit{non-existence} of leaves --- proving that a certain leaf record does \emph{not} exist under a given root $\xi$ --- which in turn leads to limited assurances for retrieval queries, including a lack of authenticated denial.

An index-ordered Merkle tree (IOMT) \cite{mohanty16, mohanty2012efficient} corrects this last deficiency of ordinary Merkle trees by treating leaf records as a virtual circularly-linked list, which facilitates proofs of non-existence while maintaining other desirable properties of Merkle trees, such as logarithmic update/verification time.

All leaf records in an IOMT are a 3-tuple consisting of the fields $(IDX, IDX^{Next}, VAL)$. $IDX$ and $IDX^{Next}$ are the index of the current leaf record and the index of next linked leaf record, respectively. $VAL$ is a fixed-length (a hash or monotonic counter) value kept as a succinct representation of a record with the index $IDX$.

A leaf record of the form $(a, a^{Next}, 0)$, where $VAL = 0$, is a special case called a \emph{placeholder}. Placeholders are used for record initialization or proving uninitialized indices.

\begin{figure}[!h]
\begin{center}
 \begin{tikzpicture}[scale=1]

\tikzstyle{every node}=[draw,rectangle,inner sep=2pt]

\draw (0.75,3) node (v34) {\footnotesize $(3,4, \omega_3)$};
\draw (2.75,3) node (v13) {\footnotesize $(1,3, \omega_1)$};
\draw (4.75,3) node (v47) {\footnotesize $(4,7, \omega_4)$};
\draw (6.75,3) node (v71) {\footnotesize $(7,1, \omega_7)$};

\tikzstyle{every node}=[draw,circle,inner sep=1pt]

\draw (0.75, 4) node (v0) {\footnotesize $v_{0}$};
\draw (2.75, 4) node (v1) {\footnotesize $v_{1}$};
\draw (4.75, 4) node (v2) {\footnotesize $v_{2}$};
\draw (6.75, 4) node (v3) {\footnotesize $v_{3}$};

\draw[->]  (v34) -- (v0);
\draw[->]  (v13) -- (v1);
\draw[->]  (v47) -- (v2);
\draw[->]  (v71) -- (v3);

\draw (1.75,5) node (v01) {\footnotesize $v_{01}$};
\draw (5.75,5) node (v23) {\footnotesize $v_{23}$};

\draw[->] (v0) -- (v01);
\draw[->] (v1) -- (v01);
\draw[->] (v2) -- (v23);
\draw[->] (v3) -- (v23);

\draw (3.75,6) node (v0f) {\footnotesize $\xi$};

\draw[->]  (v01) -- (v0f);
\draw[->]  (v23) -- (v0f);

\draw [ultra thick, dashed, ->] (v34) to [out=30,in=150] (v47);
\draw [ultra thick, dashed, ->] (v47) to [out=30,in=150] (v71);
\draw [ultra thick, dashed, ->] (v71.south) to [out=-150,in=-30] (v13.south);
\draw [ultra thick, dashed, ->] (v13.south) to [out=-150,in=-30] (v34.south);

\end{tikzpicture}
\end{center}
\caption{An index-ordered Merkle tree (IOMT) with 4 leaves ($h=2$).}
\label{fig:iomtree}
\end{figure}
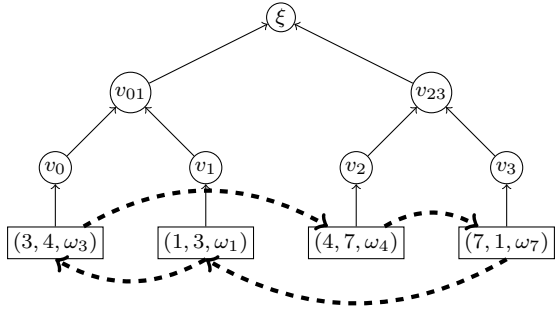

Integrity of the IOMT is maintained by requiring that $IDX < IDX^{Next}$ for all leaf records, except for the leaf record with the greatest index $IDX_{max}$, in which case $IDX^{Next} = IDX_{min}$, ensuring the circular linkage of the virtual list. For a tree with only one leaf record, $IDX_{max} = IDX_{min}$, so $IDX = IDX^{Next}$ for the single record in the tree.

Figure \ref{fig:iomtree} shows an example IOMT consisting of four leaves with a height $h=2$. The leaf records of the IOMT are linked in the order $(3,4, \omega_3) \rightarrow (4,7, \omega_4) \rightarrow (7,1, \omega_7) \rightarrow (1,3, \omega_1) \rightarrow (3,4, \omega_3)$. Note that the ordering of leaf records within the tree is insignificant; only the virtual ordering by their indices matters.

In an IOMT, a leaf record $(b, b^{Next}, \omega_b)$ is said to \emph{enclose} another index $a$ if and only if

\[
(b < a < b^{Next}) \lor (b^{Next} <= b < a) \lor (a < b^{Next} <= b).
\]

In the remainder of this paper, we take the more compact form ``$(b, b^{Next})$ \textbf{encloses} $a$'' to be synonymous with the above expression.

In our example tree (Figure \ref{fig:iomtree}), the existence of the encloser leaf $(4,7, \omega_4)$ proves that no leaf record exists with an index between $4$ and $7$; in other words, proving that the leaf record $(4, 7, \omega_4)$ exists in an IOMT with root $\xi$ implicitly proves the \emph{non-existence} of leaf records with the indices enclosed by $(4, 7)$ in the same tree. More generally, encloser leaves can be used to prove that no leaf with a certain index exists in an IOMT; i.e. the existence of a leaf $(b, b^{Next}, \omega_b)$ in an IOMT implies the non-existence of any leaf records with indices enclosed by $(b, b^{Next})$. The model of encloser leaves and their ability to prove leaf non-existence is useful for generating \emph{authenticated denial} responses in the event that a record requested by a user does not exist.

\section{Related Work}
\label{related}
One of the earliest works in the field of maintaining a data repository's integrity is the Secure Untrusted Data Repository, or SUNDR \cite{li2004secure}. The authors describe the property of ``fork consistency'' to detect integrity and consistency issues in data stored securely on untrusted servers. SUNDR focuses heavily on maximizing the concurrency of individual operations, and increasing the efficiency of the system as a whole. The SUNDR model does not include a trusted entity, which leads to some inherent limitations; as a result, SUNDR enables an attack known as \emph{forking}, in which it is possible for a malicious server to deceive two users into seeing separate and inconsistent views of the same repository.

Ensuring security assurances for cloud or remote data storage services using authenticated data-structure has been a widely studied domain. Erway et. al. \cite{erway2015dynamic} have proposed the use of a authenticated dictionary based on rank information to prove data possession. Privacy preservation of stored data using audit logs \cite{wang2010privacy, wang2014oruta} and cryptographic models \cite{kamara2010cryptographic} have been used to provide confidentiality assurances. In the work done by Bowers et.al. \cite{bowers2009hail}, cryptographic models in a distributed system were used and evaluated to provide ``proofs of retrievability'' towards availability and integrity of stored data. Access control mechanisms have also been explored by various authors \cite{wang2010hierarchical, wan2012hasbe} to enable sharing of private data across remote servers. Several other security challenges (such as search, range query, security overhead, etc.) in such cloud based environment have also been identified by Ren et.al. \cite{ren2012security}.

One such approach is the use of the authenticated data structure --- Merkle trees \cite{merkle88}, which been applied to a wide range of applications to ensure integrity and trust of data publication  \cite{devanbu2002authentic}, authentication schemes \cite{li2014efficient, buchmann2008merkle}, tamper evident logging \cite{crosby2009efficient}, database integrity \cite{martel2004general, li2006dynamic, mykletun2006authentication}, and routing \cite{hu2003sead}, among others. In the study done by Sarmenta et al. \cite{sarmenta06}, Merkle trees, in conjunction with a trusted platform module (TPM), were used for creating virtual monotonic counters for count-limited objects. These objects can then be used to provide update/utilization assurance for virtual payments, data storage, encryption/decryption keys, etc.

A similar approach was proposed by Tate et al. \cite{tate13}, in which the use of a TPM in a system for providing distributed data storage to multiple users was developed and evaluated. Similar to \cite{sarmenta06}, it relies on the use of a ordinary Merkle tree (or hash tree) to maintain a collection of virtual monotonic counters, with the root of the tree being protected the TPM. While both approaches are able to provide assurances of integrity due to the use of a Merkle tree, they are unable to provide the desirable feature of \emph{authenticated denial} --- proofs that certain data does \emph{not} exist within the repository.

In order to address the limitations above, an IOMT-based approach with a trusted boundary for root storage was proposed in the prior work done by Mohanty et al. \cite{mohanty14, mohanty11}. The paper outlines a theoretical system known as ``Cloud Storage Assurance Architecture'' (CSAA), which uses IOMT-based virtual monotonic counters and self-memoranda to address the issue of secure cloud file storage. This paper is an extension of this previous work; the IOMT authenticated data structure has been modified for use with container images, $DockerFile$, and $ComposeFile$, resulting in the ``Trustworthy Container Repository'' (TCR) infrastructure. We present a preliminary version of TCR for experimental evaluation of the performance characteristics of the proposed model, and compare it with a similar (simulated) non-secure container repository.

A comparative container repository to the proposed TCR architecture is the Docker Hub \cite{dockerhub}. It is one of the most widely used centralized repository for hosting docker $images$. Developers frequently reuse other pre-built container images to avoid building an image from scratch, and any user can create an account to host their own container images. While the repository contains, pre-built $images$, the repository contains no information about the build code  $Dockerfile$ and the stack file $Composefile$. The trust in the container images is based on the reputation of the user or the developer community responsible for building it. As a result, a study conducted by Gummaraju et al. \cite{gummaraju2015over} show almost 30\% of the container images hosted on Docker Hub contain vulnerabilities which make them highly susceptible to security attacks. Similar study of security vulnerabilities by Shu et al. \cite{shu2017study}, found even the trusted official and community based repositories contain more than 180 vulnerabilities on average. 

Most recently, in a 2018 security incident \cite{goodin18} a malicious user (with a legitimately created account) pushed several images masquerading as database servers, with cryptocurrency-mining malware injected, to Docker Hub. Although this incident did not involve a compromise of Docker Hub itself, the malicious images involved were still pulled several million times before being taken down. While this incident was limited to only 17 container images, a compromise of Docker Hub, which could allow the backdooring of \emph{every} image stored in the repository, could be far worse. 

Docker Hub includes aims to mitigate some of the security issues by using a ``Content Trust'' \cite{contenttrust} mechanism to ensure all $images$ the Docker client works with have been signed by a trusted publisher, and are the most recent/freshest version available. Within the system, each publisher (of container images) holds a root key, termed the ``Offline key'', and several ``Tagging keys'' (signed by the ``Offline key'), one for each container image \cite{contenttrust2}. Whenever an image is first retrieved/``pulled'' from the hub, the Docker client remembers the public key associated with the image's publisher, and uses it to authenticate all future connections to the hub (similar to host authentication used in the SSH protocol). All image content is signed with the image-specific Tagging key \cite{contenttrust3}.

Freshness of container data is ensured by a system called  ``The Update Framework'' (TUF) \cite{cappos10}. TUF relies on a metadata file that is periodically signed by a ``Timestamp key'' (in turn signed by the Offline key). It is the responsibility of the users to request the servers to periodically poll the repository server to check for new updates and obtain the most recent images. 

While the security of  ``Content Trust'' provision certain security measures for collaborative container development, however, its utility is limited by its opt-in nature (integrity checking is disabled by default) and complex key management system, which requires periodic re-signing with a ``Timestamp key''. It also fails to address the transparency needed to provide trust in the repository service. More specifically, the trust in the service operation is not assured with malicious entities at service administrative level capable of performing unwarranted modifications on the container images. $Dockerfile$ build files are not tracked within the service, which provide explicit information of the content within the containers and can reduce container vulnerability by ensuring the upto data software is used in its creation. Similarly, verification of $Composefile$ can enable proper utilization of container images in deployment. Additionally, Content Trust cannot provide authenticated denials, making improper denial-of-service a possibility.

\section{Trustworthy Container Repository}
\label{tcroverview}

We present a Trustworthy Container Repository (TCR) infrastructure that facilitates the secure storage of software container images, such as those used by Docker, by leveraging the utility of IOMTs and the trustworthiness of a trusted module. The TCR infrastructure consists of three primary entities: 1) a trusted module --- \textbf{T}; 2) an untrusted service provider --- \textbf{\textit{S}}; and 3) any number of participating users --- \textbf{\textit{U}}.

The TCR infrastructure provides the following assurances, based on the security assumptions that 1) the module \textbf{T} is trustworthy; 2) the hash function chosen as $h()$ is preimage-resistant; and 3) there is secure, verifiable communication between entities (based on prior secret-sharing):

\begin{itemize}
    \item Integrity
    \begin{itemize}
        \item[\textbf{I1}] - \textbf{\textit{S}} cannot pass off tampered container images as legitimate.
        \item[\textbf{I2}] - \textbf{\textit{S}} cannot pass off tampered build code (Dockerfile) as legitimate.
        \item[\textbf{I3}] - \textbf{\textit{S}} cannot pass off tampered deployment code (Docker Compose file) as legitimate.
        \item[\textbf{I4}] - Only \textbf{\textit{U}} with sufficient access privileges ($U_{acc} >= 2$) can modify a container.
    \end{itemize}
    \item Availability
    \begin{itemize}
        \item[\textbf{A1}] - \textbf{\textit{S}} cannot deny existence of container records if they exist (authenticated denial).
        \item[\textbf{A2}] - \textbf{\textit{S}} cannot deny existence of container versions if they exist.
    \end{itemize}
    \item Confidentiality
    \begin{itemize}
        \item[\textbf{C1}] - \textbf{\textit{S}} cannot view the contents of sensitive containers.
        \item[\textbf{C2}] - \textbf{\textit{S}} cannot modify ACLs without an authorized request from a user with sufficient permissions ($U_{acc} = 3$).
    \end{itemize}
    \item Consistency
    \begin{itemize}
        \item[\textbf{F1}] - \textbf{\textit{S}} cannot deceive users into seeing inconsistent views of the same repository.
    \end{itemize}
\end{itemize}

\subsection{TCR Entities and Security Model}

The role of the service provider \textbf{\textit{S}} is to maintain an authenticated data structure (ADS), handle user requests, and communicate with the trusted module \textbf{T}. \textbf{\textit{S}} can be comprised of any server-grade hardware, with no resource constraints on its capabilities.

The service provider \textbf{\textit{S}} is assumed to be completely untrusted. It can tamper with user data, improperly deny service, and share encrypted container images with unauthorized users. However, the TCR infrastructure ensures that such misbehavior by \textbf{\textit{S}} is either detectable by users, or that it cannot reveal sensitive information; e.g. if \textbf{\textit{S}} claims that a container does not exist, it will be unable to produce the proof of trust from \textbf{T} that the requesting user expects when denied service, alerting the user to the misbehavior. Also, if a user chooses to encrypt a sensitive container, \textbf{\textit{S}} will be unable to learn anything from the image it is given, because it is protected by encryption.

The module \textbf{T} is the only trusted entity in the TCR infrastructure. \textbf{T} consists of tamper-resistant, non-volatile memory and a cryptographic processor. The non-volatile memory is used to store a copy of the container IOMT root $\xi$, user keys (shared secrets between users and the module), and the module secret $\chi$, which is a secret value randomly generated upon module initialization. Any tampering with \textbf{T} will lead to the immediate erasure of all sensitive values, leaving \textbf{\textit{S}} unable to provide proofs of trust. Although the root $\xi$ is a public value (\textbf{\textit{S}} can compute it), the copy of $\xi$ stored in \textbf{T} is protected, and can only be modified by the cryptographic procedures executed inside \textbf{T}.

The cryptographic processor in \textbf{T} can be used to generate \emph{certificates}, a form of self-memoranda (see Section \ref{certificates}). The processor also has the capability to execute simple cryptographic procedures: Transformation Procedures (updates to the module IOMT root) and Integrity Verification Procedures (functionality for providing authenticated proofs of container integrity to \textbf{U}).

TCR does not prescribe exactly how \textbf{T} is to be implemented. It could be a resource limited hardware module (similar to a Trusted Platform Module \cite{morris2011trusted}), or something else entirely --- the implementation details are outside the scope of this paper. We simply assume that \textbf{T} provides the necessary functionality (security and trust).

The trustworthiness of the module \textbf{T} is crucial for the security of TCR infrastructure as a whole. Therefore, it is desirable to minimize the required functionality of \textbf{T}, easing verification that the module is free from any undesired functionality (i.e. a backdoor). Additionally, a small feature set allows for better shielding from physical introspection, and in turn, increased security for the sensitive data stored in the module. To this end, TCR only requires that \textbf{T} perform a fixed set of relatively simple operations: the protected storage of small amounts of data, and simple cryptography based on a hash function, $h()$.

\subsection{Data Structures}
\label{datastructures}

\begin{figure}[ht!]
  \centering
  \scalebox{1}{
  \includegraphics[height=3.5in]{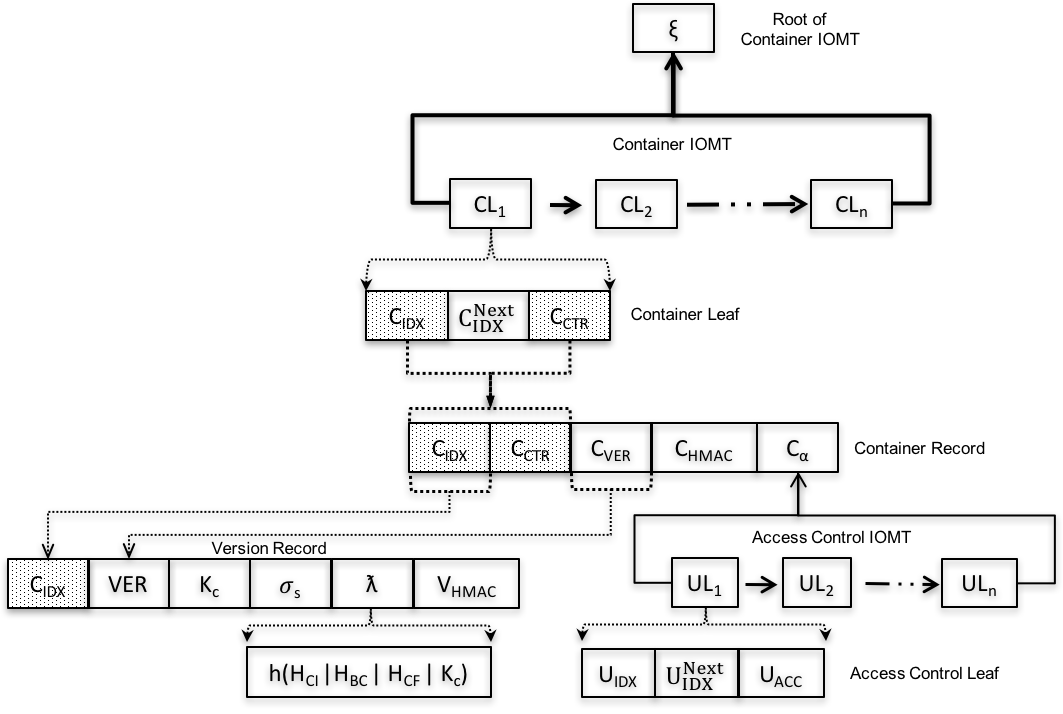}}
  \caption{Overview of TCR data structures}
  \label{fig:ds1}
  \vspace{-1em}
\end{figure}

The TCR infrastructure uses an authenticated data structure (ADS), based on IOMTs, in order to provide the aforementioned assurances. Figure \ref{fig:ds1} shows the design of the data structure, which consists of four major components --- 1) the Container IOMT; 2) the Container Record table; 3) the Container Version Table; and 4) Container Access Control IOMTs. All four components of the ADS are mapped, directly or indirectly, to the root of the Container IOMT (denoted $\xi$), which is stored within trusted module \textbf{T}, and can only be modified by TCR algorithms (described in later sections).

$\xi$ is the root of the container IOMT, in which each leaf record of the IOMT represents a single container identified by unique index. The leaf record also contains the next index the leaf is linked to, and an update counter tracking any updates that where performed on the record (and in turn, on the container). The leaf record of the container IOMT is of the following form:

\begin{equation}
    CL_i = <IDX, IDX^{Next}_{IDX}, CTR>,
\end{equation}

where $IDX$ = Container Index; $IDX^{Next}_{IDX}$ = Next Container Index; $CTR$ = Container Counter. The hash value ($h(CL_i)$) representative of the leaf record is kept as the record's corresponding leaf node in the container IOMT ($1 \leq i \leq n$; $n$ is the maximum number of leaves in the IOMT).

The container index ($IDX$) acts as the index (primary key) for container records stored in a SQL database. Each record is of the form 

\begin{equation}
CR_i = <C_{IDX}, C_{CTR}, C_{VER}, C_{HMAC}, C_\alpha>,
\end{equation}

where $C_{IDX}$ and $C_{CTR}$ are from the container leaf record; $C_{VER}$ is a counter for the number of versions of the container; $C_{HMAC} = HMAC([C_{IDX}, C_{CTR}, C_{VER}, C_\alpha], \chi)$ is a self-certificate issued by \textbf{T}, verifying the authenticity of the container record; $C_\alpha$ is the root of the corresponding container access control IOMT.

Every container has a corresponding access control IOMT, which contains an access level for each collaborating user for a container. The leaf records of the access control IOMT are of the form

\begin{equation}
    UL_j = <U_{IDX}, U^{Next}_{IDX}, a>,
\end{equation}

where $U_{IDX}$ is the user index; $U^{Next}_{IDX}$ is the next user index; and $a \in \{0,1,2,3\}$ is the access level of the user, with $0$ = no access, $1$ = read-only, $2$ = read/write, and $3$ = read/write + ACL modify.

A newly initialized container (with the index $C_{IDX}$) has $C_{VER} = 0$, meaning that it has no version history; all other containers have a value of $C_{VER} >= 1$, implying the existence of all versions $1 <= VER <= C_{VER}$. Each version reflects an update made to the contents of the container, and is described by a \emph{version record}. The service provider \textbf{\textit{S}} maintains such records in a database with the following form:

\begin{equation}
VR_j = <C_{IDX}, VER, \mu_{cs}, \sigma_s, \lambda, V_{HMAC}>,
\end{equation}

$C_{IDX}$ is the container index; $VER$ is the version number $(1 <= VER <= C_{VER})$; $\mu_{cs} = h(\sigma \parallel C_{IDX})$ is a commitment to the container encryption secret $\sigma$ and the index $C_{IDX}$; $\sigma_s$ is the encrypted container secret; $\lambda = h(H_{CI} \parallel H_{BC} \parallel H_{CF} \parallel \mu_{cs})$ is a commitment to the container image hash ($H_{CI}$), the container build code hash ($H_{BC}$), the container configuration file ($H_{CF}$), and the container encryption commitment ($\mu_{cs}$); $V_{HMAC} = HMAC([C_{IDX}, VER, \lambda], \chi)$ is the module self-certificate (issued by \textbf{T}) for the version record.

In addition to the four data structures listed above, the service provider \textbf{\textit{S}} is expected to store the contents of the container image files (TAR archives in the case of Docker), associated build code (Dockerfiles), and configuration (Docker Compose files).

\subsection{TCR Certificates}
\label{certificates}
One of the key capabilities of the trusted module \textbf{T} is its ability to issue \emph{certificates}, a form of self-memoranda. Certificates serve to either prove some fact about an IOMT, or prove that a record with certain values exists. A certificate consists of two parts --- 1) the memorandum, whose contents $Cert$ are dependent on the certificate type, and 2) a hash message authentication code (HMAC) $\rho$, computed as:

\[
\rho = HMAC(Cert, \chi).
\]

where $Cert = [type, [v_1, v_2, \cdots, v_n]]$ with $v_1 \cdots v_n$ being the values in the certificate and $type$ being the type of certificate issued; $\chi$ the module secret.

We will use the notation $Cert.v1$ to refer to specific fields of a certificate. We will also use the representation $X \rightarrow X'$, suggesting an old node $X$ update to new node $X'$, and similarly $Y \rightarrow Y'$, representing an old root of IOMT $Y$ update to new root $Y'$.

The TCR infrastructure requires \textbf{T} to generate six different types of certificates. Each type of certificate either proves some fact about an IOMT (a root-node mapping or node-record update) or authenticates a container or version record.

\begin{itemize}
\item \textbf{Node Update (NU)}
\begin{description}
    \item[Input]  ($X$ - Old node, $X'$ - New node, $[X_{comp}]$ - List of complementary nodes to $X$)
    \item[Output] $\rho_{NU} = HMAC([type = NU, [X, Y, X', Y']], \chi)$ 
    \item[Description] NU certificates, issued by $F_{nu}()$ (Algorithm~\ref{alg:nu}), verify that some IOMT node $X$ is a child node of an IOMT with root $Y$, and the transformation $X \rightarrow X'$ will result in a change of IOMT root $Y \rightarrow {Y}'$. The procedure uses the Merkle tree calculation procedure $F_{mt}()$ (Algorithm~\ref{alg:merkle_root}) to calculate the root node values, using the set of complementary nodes given by $[X_{comp}]$.
\end{description}

\begin{algorithm}
\caption{Node Update (NU) certificate generation procedure}\label{f_nu}
\label{alg:nu}
\begin{algorithmic}[1]
\Procedure{$F_{nu}$}{$X, {X}', [X_{comp}]$}
\State{$Y \gets f_{mt}(X, [X_{comp}])$} \Comment {Calculate old root}
\If {$X = {X}'$}
\State {${Y}' \gets Y$} \Comment{if $X = {X}'$, then $Y = Y'$}
\Else
\State {${Y}' \gets f_{mt}({X}', [X_{comp}])$} \Comment {Update to new root ${Y}'$}
\EndIf
\State {$Cert_{NU} \gets [NU, [X,Y,{X}',{Y}']]$}
\State \Return {$\rho_{NU} \gets HMAC(Cert, \chi)$} \Comment{Sign with module secret $\chi$}
\EndProcedure
\end{algorithmic}
\end{algorithm}

\item \textbf{Record Verify (RV)}

\begin{description}
\item[Input] ($\rho_{NU}, Cert_{NU}, <IDX, IDX^{Next}, VAL>, IDX'$)
\item[Output] $\rho_{RV} = HMAC([type = RV, [IDX, VAL, Y]], \chi)$ 
\item[Description] The RV certificate procedure (Algorithm~\ref{alg:rv}) maps the value in a node to its IOMT root. Using a NU certificate of the form $Cert_{NU} = [NU, [X,Y,{X}',{Y}']]$ (where $X=X'$ and $Y=Y'$), it maps the values $IDX, VAL$ to IOMT root $Y$.

Optionally, an index $IDX'$ such that $(IDX, IDX^{Next})$ encloses $IDX'$ can be passed to the function. In this case, a second certificate $Cert_{RV2}$ is generated of the form $[RV, [IDX', 0, Y]]$, proving no leaf record with index $IDX'$ exists within the IOMT with root $Y$.
\end{description}

\begin{algorithm}
\caption{Record Verify (RV) certificate generation procedure}\label{f_rv}
\label{alg:rv}
\begin{algorithmic}[1]
\Procedure{$F_{rv}$}{$\rho_{NU}, Cert_{NU}, <IDX, IDX^{Next}, VAL>, IDX'$}
\State {$X \gets h(IDX \parallel IDX^{Next} \parallel VAL)$}
\If {$Cert_{NU}.X \neq X$}
\Return NULL
\EndIf
\State {$Y \gets Cert_{NU}.Y$}
\State {$Cert_{RV1} \gets [RV, [IDX, VAL, Y]]$}
\If {$(IDX, IDX^{Next})$ \textbf{encloses} $IDX'$} \Comment {Enclosure verification.}
\State {$Cert_{RV2} \gets [RV, [IDX', 0, Y]]$} \Comment {No node with index $IDX'$ exists under root $Y$.}
\State \Return {$\rho_{RV} \gets HMAC((Cert_{RV1}, Cert_{RV2}), \chi)$} 
\Else
\State \Return {$\rho_{RV} \gets HMAC(Cert_{RV1}, \chi)$} 
\EndIf
\EndProcedure
\end{algorithmic}
\end{algorithm}

\item \textbf{Record Update (RU)}

\begin{description}
\item[Input] ($\rho_{NU}, Cert_{nu}, <IDX, IDX^{Next}, VAL>, VAL'$)
\item[Output] $\rho_{RU} = HMAC([type = RU, [IDX, VAL, Y, VAL', Y']], \chi)$ 
\item[Description] A RU certificate (Algorithm~\ref{alg:ru}) describes the effect that changing the value field in an IOMT leaf $<IDX, IDX^{Next}, VAL>$ from $VAL \rightarrow VAL'$ has on the root $Y$ ($Y \rightarrow {Y}'$).

Using a node update certificate ($Cert_{NU}$ and $\rho_{NU}$) the procedure $F_{ru}()$ (Algorithm~\ref{alg:ru}) verifies that $X = h(IDX \parallel IDX^{Next} \parallel VAL)$ and ${X}' = h(IDX \parallel IDX^{Next} \parallel VAL')$, before returning a certificate of the form $Cert_{RU} = [type = RU, [IDX, VAL, Y, VAL', Y']]$.
\end{description}

\begin{algorithm}
\caption{Record Update (RU) certificate generation procedure}\label{f_ru}
\label{alg:ru}
\begin{algorithmic}[1]
\Procedure{$F_{ru}$}{$\rho_{NU}, Cert_{nu}, <IDX, IDX^{Next}, VAL>, VAL'$}
\State {$X \gets h(IDX \parallel IDX^{Next} \parallel VAL)$}
\State {$X' \gets h(IDX \parallel IDX^{Next} \parallel VAL')$}
\If {($Cert_{NU}.X \neq X \lor Cert_{nu}.X' \neq X'$)}
\Return NULL
\EndIf
\State {$Y \gets Cert_{NU}.Y$}
\State {$Y' \gets Cert_{NU}.Y$}
\State {$Cert_{RU} \gets [RU, [IDX, VAL, Y, VAL', Y']]$}
\State \Return {$\rho_{RU} \gets HMAC(Cert_{RU}, \chi)$} 
\EndProcedure
\end{algorithmic}
\end{algorithm}

\item \textbf{Root Equivalence (EQ)}

\begin{description}
\item[Input]  ($Cert_{NU1}, \rho_{NU1}, Cert_{NU2}, \rho_{NU2}, <IDX, IDX^{Next}, VAL_{IDX}>, IDX'$)
\item[Output] $\rho_{EQ} = HMAC([type = EQ, [Y, Y'']], \chi)$ 
\item[Description] An EQ certificate (Algorithm~\ref{alg:eq}) verifies that two IOMT root values $Y$ and $Y''$ are equivalent roots --- $Y''$ contains only an additional placeholder (a leaf with a value $VAL = 0$) inserted into the tree.

Inserting a placeholder with index $IDX'$ into an IOMT with root $Y$ requires that there be an encloser leaf $<IDX, IDX', VAL_{IDX}>$, such that $(IDX, IDX^{Next})$ encloses $IDX'$ (see Section \ref{iomt}). Given such a leaf, it takes two updates to the tree to insert a placeholder: the first update changes the value of the $IDX^{Next}$ field of the encloser leaf to $IDX'$ (such that the first leaf is then of the form $<IDX, IDX', VAL_{IDX}>$). Then, the placeholder node $<IDX', IDX^{Next}, 0>$ is inserted, linked to the original $IDX^{Next}$ to maintain the the integrity of the linked list.

These updates can be verified using two node update certificates --- 1) $Cert_{NU1} = [NU, [X_1, Y_1, X_1', Y_1']]$, where $X_1 = h(IDX \parallel IDX^{Next} \parallel VAL_{IDX})$ and $X_1'= h(IDX \parallel IDX' \parallel VAL_{IDX})$; and 2) $Cert_{NU2} = [NU, [X_2, Y_2, X_2', Y_2']]$, where $X_2=0$ and $X_2'= h(IDX' \parallel IDX^{Next} \parallel 0)$.

Given these two certificates, the module \textbf{T} can infer that $Y$ and $Y''$ are equivalent roots, with $Y''$ having an additional placeholder node with index $IDX'$ inserted. The module can then issue a certificate of the form $[EQ, [Y, Y'']]$. 
\end{description}

\begin{algorithm}
\caption{Root Equivalence (EQ) certificate generation procedure}\label{f_eq}
\label{alg:eq}
\begin{algorithmic}[1]
\Procedure{$F_{eq}$}{$Cert_{NU1}, \rho_{NU1}, Cert_{NU2}, \rho_{NU2}, <IDX, IDX^{Next}, VAL_{IDX}>, IDX'$}
\If {$\lnot((IDX, IDX^{Next})$ \textbf{encloses} $IDX')$}
\State \Return $Cert_{NULL}$ \Comment{Not an encloser leaf}
\EndIf
\State {$X_1 = h(IDX \parallel IDX^{Next} \parallel VAL_{IDX})$} \Comment{Hash of old and new encloser leaf}
\State {$X_1'= h(IDX \parallel IDX' \parallel VAL_{IDX})$}
\State {$X_2 \gets 0$} \Comment{Initially no placeholder}
\State {$X_2'= h(IDX' \parallel IDX^{Next} \parallel 0)$}

\If {$Cert_{NU1}.X_1 \neq X_1 \lor Cert_{NU1}.X_1' \neq X_1' \lor\newline\hspace*{2.6em}
      Cert_{NU2}.X_2 \neq X_2 \lor Cert_{NU2}.X_2' \neq X_2'$}
\State \Return NULL \Comment {Certificate node values do not match.}
\EndIf

\If{$Cert_{NU1}.Y_1' \neq Cert_{NU2}.Y_2$}
\Return NULL \Comment {Certificates do no form a chain.}
\EndIf

\State {$Y \gets Cert_{NU1}.Y_1$}
\State {$Y'' \gets Cert_{NU2}.Y_2'$}

\State {$Cert_{EQ} \gets [EQ, [Y, Y'']]$}
\State \Return {$\rho_{RU} \gets HMAC(Cert_{EQ}, \chi)$}
\EndProcedure
\end{algorithmic}
\end{algorithm}

    \item \textbf{Container Record (CR)}
    \begin{description}
    \item[Input]  ($Cert_{RV}, \rho_{RV}, Cert_{RU}, \rho_{RU}, Cert_{CR}, \rho_{CR}, IDX, \mu, C_{CTR}, C_{VER}, v=C_{alpha}$)
    \item[Output] $\rho_{CR} = HMAC([CR, [IDX, C_{CTR}, C_{VER}, C_{\alpha}]], \chi)$ 
    \item[Description] A container certificate is generated to map the container record to the root of the IOMT $\xi$. Container records are created by $C_{tp}$ procedure (Algorithm \ref{alg:tp}) based on an user request. The resulting updates to the container record of form [$IDX, \mu, C_{CTR}, C_{VER}, C_{alpha}$]
    \end{description}
    
A CR certificate has the fields $[index, counter, version, \alpha]$.

    \item \textbf{Version Record (VR)}
    
   \begin{description}
    \item[Input]  ($Cert_{RV}, \rho_{RV}, Cert_{RU}, \rho_{RU}, Cert_{CR}, \rho_{CR}, IDX, \mu, C_{CTR}, C_{VER}, v$)
    \item[Output] $\rho_{VR} = HMAC([VR, [IDX, VER, \lambda]], \chi)$ 
    \item[Description] Version record verifies a version ($C_{VER}$) of a container along with its index ($IDX$) and container hash/secret commitment $\lambda$ to the root of the IOMT. Similar to the container record $C_{tp}$ procedure (Algorithm \ref{alg:tp}) creates the version record based on a user request. 
    \end{description}
    
\end{itemize}

\subsection {TCR Procedures}
TCR procedures use the generated certificates (and generate CR and VR certificates) to update the stored root in the module (reflecting the changes to the datastructure), retrieve verified information, and ensure proper storage / retrieval of container secrets. These operations are based on user \textbf{\textit{U}} requests to the service provider \textbf{\textit{S}}. \textbf{\textit{S}} utilzes the capabilities of \textbf{T} to perform updates and retrieve information verified by it. The description of the procedures is as follows:

\begin{itemize}
\item \textbf{Placeholder Insertion/Deletion: $F_{ph}()$}

\begin{description}
\item[Input]  ($Cert_{EQ}, \rho_{EQ}$)
\item[Output] None
\item[Description] $F_{ph}()$ (Algorithm \ref{f_ph}) accepts an EQ certificate of the form $[[Y, Y'], EQ]$. If the current root of container IOMT $\xi = Y$ (stored by \textbf{T}), then the root $\xi$ is toggled to $Y'$. Otherwise, if $\xi = Y'$, the module root is instead changed to $Y$. The procedure is invoked by the service provider \textbf{\textit{S}} in order to insert a placeholder into the container IOMT, and does not require any form of authentication, i.e. anyone with access to the module is allowed to execute it, since it does not affect the contents of the container repository.
\end{description}

\begin{algorithm}
\caption{Placeholder insert/delete procedure}\label{f_ph}
\begin{algorithmic}[1]
\Procedure{$F_{ph}$}{$\rho_{EQ}, Cert_{EQ}$}

\If {$\xi = Cert_{EQ}.Y$}
\State {$\xi \gets Cert_{EQ}.Y'$}
\ElsIf {$\xi = Cert_{EQ}.Y'$}
\State {$\xi \gets Cert_{EQ}.Y$}
\EndIf
\EndProcedure
\end{algorithmic}
\end{algorithm}

\item \textbf{Container Operation: $F_{tp}()$}

\begin{description}
\item[Input]  ($[type, IDX, C_{CTR}, v], \mu, Cert_{RV}, \rho_{RV}, Cert_{RU}, \rho_{RU}, Cert_{CR}, \rho_{CR}$)
\item[Output] $Cert_{CR}, \rho_{CR}, Cert_{VR}, \rho_{VR}$
\item[Description] $F_{tp}()$ (Algorithm ~\ref{f_tp}) handles user requests for container creation, container updates, and ACL modification. The operations on the data structure are described by a user request of the form $[type, IDX, C_{CTR}, v]$, where $type$ distinguishes between container and ACL updates, $IDX$ is the index of the container to be modified, $C_{CTR}$ is the current value of the container counter, and $v = \lambda$ for container updates, or $v = C_{\alpha}$ for ACL IOMT updates. The request is accompanied by the user's signature, of the form

\[
\mu = HMAC([type, IDX, C_{CTR}, v], K_i),
\]

where $K_i$ is the shared secret between the user and \textbf{T}.

Container creation is treated as an ACL update with $C_{CTR} = 0$. In this case, the input to the $F_{tp}()$ consists of only a $RU$ certificate of the form $Cert_{RU} = [RU, [IDX, 0, Y = \xi, 1, Y' = \xi']]$. The certificate verifies that no container with index $IDX$ exists in the repository with root $Y = \xi$, and that the root must be updated from $Y \rightarrow Y'$ to reflect inserting a container leaf with $C_{CTR} = 1$. Successful completion of the operation issues a certificate of form $Cert_{CR} = [[IDX, C_{CTR}=1, C_{VER}=0, C_{\alpha} = v], CR]$. \textbf{T} will also update its internal IOMT root $\xi \rightarrow \xi'$.

For container updates and ACL updates, three certificates are required as input --- 1) a RU certificate of the form $Cert_{RU} = [RU, [IDX = C_{IDX}, VAL=C_{CTR}, Y=\xi, VAL'=C_{CTR} + 1, Y'=\xi']]$; 2) a CR certificate of the form $[[IDX, C_{CTR}, C_{VER}, C_{\alpha}], CR]$; and 3) a RV certificate of the form $[RV, [IDX = U_{IDX}, VAL=U_{ACC}, Y=C_{\alpha}]]$. The module will use the value $U_{ACC}$ to determine whether the user is allowed to execute the request (it requires $U_{ACC} >= 2$ for a container update (read/write access), or $U_{ACC} = 3$ for an ACL update). All successful requests return a new CR certificate reflecting the incremented counter value $C_{CTR}'$, and \textbf{T} updates its internal IOMT root $\xi \rightarrow \xi'$.

The CR certificate returned by a \emph{container} update will have the value $C_{VER} = C_{VER} + 1$. For an \emph{ACL} update, the certificate will have an altered $C_{\alpha}$ value but the version counter $C_{VER}$ will remain unchanged. \textbf{T} also creates a VR certificate of the form $Cert_{VR} = [VR, [IDX, Cert.C_{VER} + 1, C_{\lambda}]]$ for container updates. The service \textbf{\textit{S}} is responsible for storing all returned certificates and their corresponding MACs.

\textbf{T} acknowledges all successful requests with the value

\[
\mu_{ack} = HMAC([type, IDX, C_{CTR}, v], K_i),
\]

which is conveyed to the requesting user to prove request completion. Unsuccessful requests are \emph{not} acknowledged by the module. Instead, the user must use the output of $F_{verify}()$ determine why the request was denied.

\end{description}

\begin{algorithm}
\caption{Container repository modification procedure}\label{f_tp}
\label{alg:tp}
\begin{algorithmic}[1]
\Procedure{$F_{tp}$}{$[type, IDX, C_{CTR}, v], \mu, Cert_{RV}, \rho_{RV}, Cert_{RU}, \rho_{RU}, Cert_{CR}, \rho_{CR}$} 


\If {$Cert_{RU}.X + 1 \neq Cert_{RU}.X'$}
\Return NULL \Comment {$Cert_{RU}$ does not reflect incrementing counter}
\EndIf

\If {$Cert_{RU}.Y \neq \xi$}
\Return NULL \Comment {Current root does not match.}
\EndIf

\State {$\mu_{ack} \gets HMAC([type, IDX, C_{CTR}, v, 0], K_i)$} \Comment {Successful request response}

\If {$type = ACL \land C_{CTR} = 0$} \Comment{Container Create.}
\State {$\xi \gets Cert_{RU}.Y'$}
\State {$Cert_{CR} \gets [CR, [IDX, 1, 0, v]]$}
\State \Return {$[Cert_{CR}, \rho_{CR} = HMAC(Cert_{CR}, \xi), \mu_{ack}]$}
\EndIf

\State {$C_{CTR} \gets Cert_{CR}.C_{CTR}$}

\If {$Cert_{RU}.IDX \neq Cert_{CR}.IDX \lor {Cert_{RU}.X \neq Cert_{CR}.C_{CTR}} \lor \newline
\hspace*{2.6em}{Cert_{CR}.C_{\alpha} \neq Cert_{RV}.Y} \lor {Cert_{RV}.X \neq U_{IDX}} \lor \newline
\hspace*{2.6em}{Cert_{RV}.Y \neq \xi} \lor {C_{CTR} \neq C_{CTR}'}$}
\State \Return NULL \Comment {Inconsistent certificates}
\EndIf

\State {$U_{ACC} \gets Cert_{RV}.VAL$}

\If {$type = CONTAINER \land U_{ACC} >= 2$} \Comment{Container Update.}

\State {$C_{VER} \gets Cert_{CR}.C_{VER}$}

\State {$\xi \gets Cert_{RU}.Y'$}

\State {$Cert_{CR}' \gets [CR, [IDX, C_{CTR} + 1, C_{VER} + 1, C_{\alpha} = Cert_{CR}.C_{\alpha}]]$}
\State {$Cert_{VR} \gets [VR, [IDX, C_{VER} + 1, v]]$}

\State \Return {$[Cert_{CR}', \rho_{CR} = HMAC(Cert_{CR}', \xi), Cert_{VR}, \rho_{VR} = HMAC(Cert_{VR}, \xi)], \mu_ack]$}

\ElsIf {$type = ACL \land U_{ACC} >= 3$} \Comment{ACL Update.}

\State {$Cert_{CR}' \gets [CR, [IDX, C_{CTR} + 1, C_{VER} + 1, C_{\alpha} = v]]$}

\State \Return {$[Cert_{CR}', \rho_{CR} = HMAC(Cert_{CR}', \xi)], \mu_ack]$}

\Else

\Return NULL \Comment {User has insufficient permissions}

\EndIf

\EndProcedure
\end{algorithmic}
\end{algorithm}



\item \textbf{Version Information Verification: $F_{verify}()$}

\begin{description}
\item[Input]  ($Cert_{RV1}, \rho_{RV1}, Cert_{RV2}, \rho_{RV2}, Cert_{CR}, \rho_{CR}, Cert_{VR}, \rho_{VR}, \hat{C_{VER}}, \delta$)
\item[Output] Successful Retrieve - $\{IDX, C_{CTR}, C_{VER}, \hat{C_{VER}}, C_\alpha, \lambda, \delta\}_{K_i}$ OR Authenticated Denial - $\{IDX, \delta\}_{K_i}$
\item[Description] $F_{verify}()$ (Algorithm \ref{f_verify}) allows a user $u_i$ to retrieve authenticated information on any container version. The function succeeds if the requested container exists and the user has sufficient permissions ($a >= 1$); otherwise it will issue an authenticated denial response.

The input to $F_{verify}()$ consists of certificates, the requested version number $\hat{C_{VER}}$ (specified by the requesting user), and a nonce $\delta$ to prevent replay attacks. The certificates must be provided by \textbf{\textit{S}} based off the requested container and version number. As a special case, the user can specify $\hat{C_{VER}} = 0$, which is synonymous with $\hat{C_{VER}} = C_{VER}$, where $C_{VER}$ is the maximum version number of the requested container. However, \textbf{\textit{S}} \textit{not} \textbf{T}, is expected to handle this case by treating it as if $\hat{C_{VER}} = C_{VER}$.

At a minimum, $F_{verify}()$ requires one RV certificate of the form $Cert_{RV} = [RV, [IDX, C_{CTR}, \xi]]$, where $\xi$ is the current module root. If $C_{CTR} = 0$, the module can infer that the container does not exist, and no further certificates are necessary; \textbf{T} will return an authenticated denial response of the form $\{IDX, \delta\}_{K_i}$, where $IDX$ is the index of the requested container, $\delta$ is the nonce specified in the query, and $K_i$ is the shared secret.

If $C_{CTR} \neq 0$, implying the existence of the container, then all four certificates ($Cert_{RV1}$, $Cert_{RV2}$, $Cert_{CR}$, $Cert_{VR}$) are necessary. $Cert_{CR} = [CR, [IDX, C_{CTR}, C_{VER}, C_{\alpha}]]$ indicates the latest ACL root $C_{\alpha}$ and maximum version $C_{VER}$, and must be consistent with $Cert_{RV1}$. $Cert_{RV2} = [RV, [U_{IDX}, U_{ACC}, C_{\alpha}]]$ indicates the access level $U_{ACC}$ of the user, where its root must match that specified in $Cert_{CR}$. $Cert_{VR} = [VR, [IDX, \hat{C_{VER}}, \lambda]]$ ties $\lambda$ to the requested version number and container index.

With these four certificates, \textbf{T} can conclude that the information it has been given is consistent with the module root $\xi$. Then, depending on the user's access level $U_{ACC}$, \textbf{T} will either issue an authenticated denial of the form $\{IDX, \delta\}_{K_i}$ if $U_{ACC} = 0$, or a response of the form $\{IDX, C_{CTR}, C_{VER}, \hat{C_{VER}}, C_\alpha, \lambda, \delta\}_{K_i}$, certifying the authenticity of the values associated with the container. This response from \textbf{T} convinces the requesting user that the container exists, and also indicates the $\lambda$ value for the requested version.

The authenticated denial response returned by $F_{verify}()$ when a requested container does not exist (i.e. $C_{CTR} = 0$) is identical to the response when the requested container exists, but the user has insufficient access rights ($U_{ACC} = 0$). Assuming that \textbf{\textit{S}} cooperates (i.e. it does not reveal the reason for the denial), then users are unable to distinguish between the two cases.

\end{description}

\begin{algorithm}
\caption{Container verification}\label{f_verify}
\begin{algorithmic}[1]
\Procedure{$F_{verify}$}{$Cert_{RV1}, \rho_{RV1}, Cert_{RV2}, \rho_{RV2}, Cert_{CR}, \rho_{CR}, Cert_{VR}, \rho_{VR}, \hat{C_{VER}}, \delta$}

\If {$Cert_{RV1}.Y \neq \xi \lor Cert_{CR}.C_{CTR} \neq Cert_{RV1}.C_{CTR}$}
\State \Return NULL \Comment {Invalid certificates}
\EndIf

\State {$IDX \gets Cert_{RV1}.IDX$} \Comment{Container IOMT}
\State {$C_{CTR} \gets Cert_{RV1}.C_{CTR}$}
\State {$U_{ACC} \gets Cert_{RV2}.U_{ACC}$} \Comment{ACL IOMT}

\If {$C_{CTR} = 0 \lor U_{ACC} = 0$}
\State \Return {$\{IDX, \delta\}_{K_i}$} \Comment {Authenticated denial}
\EndIf

\If {$(\hat{C_{VER}} \neq Cert_{VR}.C_{VER}) \lor (Cert_{VR}.C_{\alpha} \neq Cert_{RV2}.C_{\alpha})$}  
\State \Return NULL \Comment {Invalid Input} \EndIf

\State {$C_{CTR} \gets Cert_{CR}.C_{CTR}$}
\State {$C_{VER} \gets Cert_{VR}.C_{VER}$}
\State {$\lambda \gets Cert_{VR}.\lambda$}
\State \Return {$\{IDX, C_{CTR}, C_{VER}, \hat{C_{VER}}, C_{\alpha}, \lambda, \delta\}_{K_i}$}

\EndProcedure
\end{algorithmic}
\end{algorithm}

\item \textbf{Encryption Key Storage: $F_{st}()$}

\begin{description}
\item[Input]  ($IDX, U_{IDX}, \sigma', \mu_{cs}$)
\item[Output] $\sigma_{s}$
\item[Description] $F_{st}()$ (Algorithm~\ref{f_st}) ensures proper storage of the container secret ($\sigma$) by encrypting it with the module \textbf{T} secret $\chi$. A user relays the intended secret to the \textbf{T}, by XORing (exclusive-OR) the key with the HMAC() of container index $IDX$, its counter $C_{CTR}$, and shared secret $K_i$ as follows:

\begin{equation}
    \sigma' = \sigma \oplus HMAC([IDX, C_{CTR}], K_i).
\end{equation}

The user also conveys a value $\mu_{cs} = h(IDX \parallel \sigma)$ along with $\sigma'$ for verification. $F_{st}()$ decrypts the encrypted secret, verifies it with $\mu_{cs}$, and then re-encrypts with the module key $\xi$ and $\mu_{cs}$ for storage.
\end{description}

\begin{algorithm}
\caption{Encryption secret verification and storage}\label{f_st}
\begin{algorithmic}[1]
\Procedure{$F_{st}$}{$IDX, U_{IDX}, \sigma', \mu_{cs}$}

\State {$\sigma \gets \sigma' \oplus HMAC([IDX, C_{CTR}], K_i)$} \Comment {Decrypt from user}
\If {$\mu_{cs} \neq h(IDX \parallel \sigma)$}
\Return {NULL}   \Comment {Check integrity}
\EndIf

\State {$\sigma_s \gets \sigma \oplus h(\mu_{cs} \parallel \chi)$} \Comment {Re-encrypt using value only known to \textbf{T}}

\State \Return $\sigma_s$ \Comment {Return encrypted key for storage by \textbf{\textit{S}}}

\EndProcedure
\end{algorithmic}
\end{algorithm}

\item \textbf{Encryption Key Retrieval: $F_{rs}()$}

\begin{description}
\item[Input]  ($Cert_{RV1}, \rho_{RV1}, Cert_{RV2}, \rho_{RV2}, Cert_{CR}, \rho_{CR}, IDX, \hat{C_{CTR}}, \hat{C_{VER}}, \sigma_s, \mu_{cs}$)
\item[Output] $\sigma_{u}$
\item[Description] $F_{rs}()$ (Algorithm ~\ref{f_rs}) allows for the retrieval of encrypted secrets stored by TCR. The request is for a container with index $IDX$, container counter $\hat{C_{CTR}}$, and version counter $\hat{C_{VER}}$. The module also requires three certificates to prove that a user has sufficient access to a container --- 1) RV certificate of the form $Cert_{RV} = [RV, [IDX, C_{CTR}, \xi]]$, where $\xi$ is the current module root; 2) CR certificate of the form $Cert_{CR} = [CR, [IDX, C_{CTR}, C_{VER}, C_{\alpha}]]$; and 3) RV certificate of the form $[RV, [U_{IDX}, U_{ACC}, C_{\alpha}]]$, which indicates the user's access level $U_{ACC}$. If the user has sufficient privileges ($U_{ACC} > 0$), the module will decrypt the secret $\sigma_s$ and re-encrypt with the integrity pad ($\mu_{cs}$) known to the user (using the XOR operation, as in Algorithm \ref{f_st}).
\end{description}

\begin{algorithm}
\caption{Encryption secret retrieval}\label{f_rs}
\begin{algorithmic}[1]
\Procedure{$F_{rs}$}{$Cert_{RV1}, \rho_{RV1}, Cert_{RV2}, \rho_{RV2}, Cert_{CR}, \rho_{CR}, IDX, \hat{C_{CTR}}, \hat{C_{VER}}, \sigma_s, \mu_{cs}$}
\If {$(IDX \neq Cert_{RV1}.IDX) \lor (C_{CTR} \neq Cert_{RV1}.C_{CTR}) \lor (C_{VER} \neq Cert_{CR}.C_{VER})$}  
\State \Return NULL \Comment {Invalid Request} \EndIf

\If {$Cert_{RV2}.U_{ACC} < 1$}  \Return NULL \Comment {Improper Privileges} \EndIf

\State {$\sigma \gets \sigma_s \oplus h(\mu_{cs} \parallel \chi)$} 

\State {$\sigma \gets \sigma_s \oplus h(\mu_{cs} \parallel \chi)$} \Comment {See Algorithm \ref{f_st}}

\If {$\mu_{cs} \neq h(IDX \parallel \sigma)$} \Return NULL \Comment {Secret invalid} \EndIf

\State {$\sigma_u \gets \sigma \oplus h(\mu_{cs} \parallel K_i)$} \Comment {Encrypt for user ($\mu_{cs}$ is stored by \textbf{\textit{S}}, but $K_i$ is secret)}

\State \Return $\sigma_u$

\EndProcedure
\end{algorithmic}
\end{algorithm}

\end{itemize}

\subsection{TCR Functionality}
\label{protocols}
\textbf{\textit{S}} responsible for maintaining the data structures described in Section \ref{datastructures}: namely, 1) IOMT describing the state of the container repository, 2) access control lists (in IOMT form) for each container, and 3) the build code and compose files for each container version. Any operations performed on the data structures needs to be verified by \textbf{T}, or else functionality will not be trusted by the users \textbf{\textit{U}} of the service. For the purpose, \textbf{\textit{S}} interfaces with the module \textbf{T} to provide functionality to container repository operation such as --- 1) Creation, 2) Content update, 3) ACL modification, 4) Information retrieval, and 5) Content retrieval.

Every request received by \textbf{\textit{S}} (from its users), it provides proof of its operation via \textbf{T}, which verifies either that the request has been executed and reflected in the state of \textbf{T} (modification requests), or that the information returned by \textbf{\textit{S}} is fresh and authentic (query requests). The responses are then relayed back to the users with $HMAC()$ signed by the module using shared keys $K_i$. 

Requests related to container creation, update, and ACL modification, \textbf{\textit{S}} to use the $F_{tp}()$ (Algorithm \ref{f_tp}) interface exposed by \textbf{T} to modify the state of the container repository. $F_{tp}()$ requires that the values describing the request ($[type, IDX, C_{CTR}, C_{VER}]$) be signed by the user secret $K_i$ (unknown to \textbf{\textit{S}}). Acknowledgement to the request include the $HMAC()$ of the request re-signed using the shared user key by \textbf{T}. 

For query requests, container information retrieval and content retrieval, \textbf{\textit{S}} uses $F_{verify}()$ (Algorithm \ref{f_verify}) and $F_{rs}()$ (Algorithm \ref{f_rs}) respectively. $F_{verify}()$ is invoked by \textbf{\textit{S}} to provide module \textbf{T} verified response to container existence, state, and update status to \textbf{\textit{U}}. $F_{rs}()$ is used to return container secrets to authenticated users, as the stored information by \textbf{\textit{S}} is in encrypted format. Neither type of query request (information or content retrieval) \emph{requires} authentication from the user. 

Section \ref{results} discusses the security implications of the functionality provided by TCR in relation to the assurances provided in greater detail.

\section{Experimental Setup}
\label{implementation}

We explore the security and performance capabilities of the proposed TCR infrastructure with a proof-of-concept implementation of the using a Client-Server architecture. Within the implementation, the service provider \textbf{\textit{S}} operates a server based machine with storage - computational capabilities (for data and records) along with the the trusted module \textbf{T}. The clients in this case are users \textbf{\textit{U}} using client software \textbf{\textit{C}} to submit request to \textbf{\textit{S}} and verify its (and \textbf{T}'s) response. 

In our software implementation of TCR, the functionality of \textbf{\textit{S}} and \textbf{T} coexist as different modules within the same monolithic program. The secure storage of module \textbf{T} is emulated as a file in the server's file system. Shared secrets between the user \textbf{\textit{U}} and \textbf{T} are also pre-computed for ease of implementation. While this experimental setup is not representative of a real-world environment, where hardware-level boundaries would be present between \textbf{\textit{S}} and \textbf{T}, the implementation allows detailed characterization of the performance of the TCR infrastructure at different loads.

A key responsibility of \textbf{\textit{S}} is to store the data, records, and data structure required by the TCR infrastructure. While the data and records stored by \textbf{\textit{S}} are no different than those of a general-purpose repository server, the key difference lies in the storage of the IOMT data structure proposed by TCR. For the purpose, we use a SQL database (SQLite) backed by an on-disk file, to persistently store the container IOMT and its corresponding ACL IOMTs.

Every IOMT is initialized with a fixed height, specified by the parameter $h=log(leaves)$. The number of leaf records in the IOMT is given by:

\begin{equation}
    leaves = 2 ^ {h}.
\end{equation}

The number of nodes in the tree is thus given by:

\begin{equation}
    nodes = 2 * leaves - 1 = 2 ^ {h + 1} - 1.
\end{equation}

\begin{figure}[!ht]
\centering
\subfloat[Merkle Tree]{\label{fig:treeform}
\begin{tikzpicture}
\tikzstyle{every node}=[draw,rectangle,inner sep=2pt]

\draw (0.75,3) node (v34) {\footnotesize $l_0$};
\draw (2.75,3) node (v13) {\footnotesize $l_1$};
\draw (4.75,3) node (v47) {\footnotesize $l_2$};
\draw (6.75,3) node (v71) {\footnotesize $l_3$};

\tikzstyle{every node}=[draw,circle,inner sep=1pt]

\draw (0.75, 4) node (v0) {\footnotesize $v_{0}$};
\draw (2.75, 4) node (v1) {\footnotesize $v_{1}$};
\draw (4.75, 4) node (v2) {\footnotesize $v_{2}$};
\draw (6.75, 4) node (v3) {\footnotesize $v_{3}$};

\draw[->]  (v34) -- (v0);
\draw[->]  (v13) -- (v1);
\draw[->]  (v47) -- (v2);
\draw[->]  (v71) -- (v3);

\draw (1.75,5) node (v01) {\footnotesize $v_{01}$};
\draw (5.75,5) node (v23) {\footnotesize $v_{23}$};

\draw[->] (v0) -- (v01);
\draw[->] (v1) -- (v01);
\draw[->] (v2) -- (v23);
\draw[->] (v3) -- (v23);

\draw (3.75,6) node (v0f) {\footnotesize $\xi$};

\draw[->]  (v01) -- (v0f);
\draw[->]  (v23) -- (v0f);
\end{tikzpicture}
} \\
\subfloat[Nodes]{\label{fig:arraynodes}
  \scalebox{0.75}{
 \begin{tikzpicture}[scale=1]
 
  \node[name=data, rectangle split,
         rectangle split horizontal,
         rectangle split parts=7,
         rectangle split draw splits=true,
         minimum height=1cm,
         align=center,
         draw] (main) 
        {
        \nodepart[text width=1cm]{one} $\xi$
        \nodepart[text width=1cm]{two} $v_{01}$
        \nodepart[text width=1cm]{three} $v_{23}$
        \nodepart[text width=1cm]{four} $v_0$
        \nodepart[text width=1cm]{five} $v_1$
        \nodepart[text width=1cm]{six} $v_2$
        \nodepart[text width=1cm]{seven} $v_3$
        };
        
        \draw[->] (main.two south) to [out=-135,in=-45] (main.one south);
        \draw[->] (main.three south) to [out=-135,in=-45] (main.one south);
        \draw[->] (main.four north) to [out=135,in=45] (main.two north);
        \draw[->] (main.five north) to [out=135,in=45] (main.two north);
        \draw[->] (main.six south) to [out=-135,in=-45] (main.three south);
        \draw[->] (main.seven south) to [out=-135,in=-45] (main.three south);
 \end{tikzpicture}
 }
 }\quad\quad\quad
\subfloat[Leaves]{\label{fig:arrayleaves}

\scalebox{0.75}{
 \begin{tikzpicture}[scale=1]  
  \node[name=data, rectangle split,
         rectangle split horizontal,
         rectangle split parts=4,
         rectangle split draw splits=true,
         minimum height=1cm,
         align=center,
         draw] (leaves) at (0,-3)
        {
        \nodepart[text width=1cm]{one} $l_0$
        \nodepart[text width=1cm]{two} $l_1$
        \nodepart[text width=1cm]{three} $l_2$
        \nodepart[text width=1cm]{four} $l_3$
        };
        
        \draw[->, opacity=0] (leaves.four south) to [out=-145,in=-65] (leaves.one south); 
 \end{tikzpicture}
 }
 }

 \caption{A Merkle tree with height $h=2$ (Figure \ref{fig:treeform}), and its equivalent array representation (Figures \ref{fig:arraynodes}--\ref{fig:arrayleaves}). Arrows in Figure \ref{fig:arraynodes} point to the parent of each node.}
\label{fig:arrayformat}
 \end{figure}
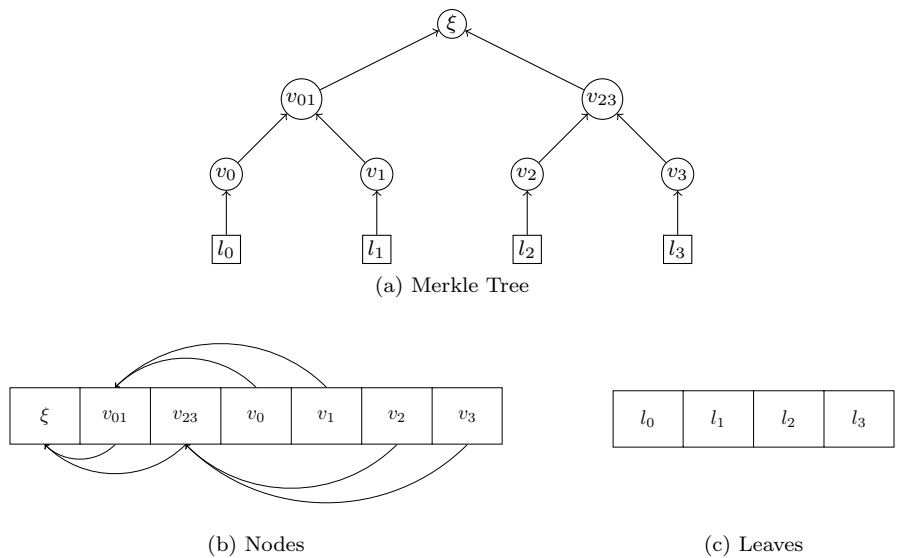

An IOMT can then be represented by sequential arrays of leaf records and nodes. Leaf records are simply numbered from left to right, with zero being the leftmost leaf record, and subsequent leaves having sequential indices within the array. Nodes are ordered in a breadth-first, left-to-right manner, starting with the root $\xi$, which is assigned the index $i_{root} = 0$. Calculations of the indices for each of the nodes is given by the following:

\begin{equation}
    i_{leftchild} = 2i_{parent} + 1
\end{equation}
\begin{equation}
    i_{rightchild} = 2i_{parent} + 2
\end{equation}
\begin{equation}
    i_{parent} = {\lfloor{i_{child} - 1} \rfloor \over 2}
\end{equation}

where the index of the left child ($i_{leftchild}$) and right child ($i_{rightchild}$) can be calculated if the index of parent ($i_{parent}$) is known. Similarly, index of parent can be calculated if the index of one of their children ($i_{child}$) is known. Figure \ref{fig:arrayformat} shows an example representation of the array format for $nodes$ and $leaves$ representation for an IOMT of height $h=3$.

The array is used for in-memory representation towards fast population of the IOMT data structure in our implementation. The in-memory representation of IOMTs uses a fixed amount of space for every value of $logleaves$. The \textbf{\textit{S}} allocates two arrays for each IOMT --- one of size $leaves$ for storing the IOMT leaves, and the other of size $nodes$ for storing the values of the IOMT nodes. The total memory used is therefore:

\begin{equation}
    sizeof(iomt leaf) * 2 ^ {logleaves} + sizeof(hash) * (2 ^ {logleaves + 1} - 1)
\end{equation}

For a more persistent representation, a SQLite database is used to store the array. Although the representation consumes large amount of space compared to the in-memory layout, the space requirement was reduced by only storing the indices for non-zero values, i.e. nodes which are not found in the database are assumed to have value zero. The implementation presents its advantages in the case of sparse trees such as recently initialized IOMT, where most nodes have value 0.

Performance evaluation of the TCR architecture is done by recording time to completion of various procedure calls for $logleaves$ value of $h = [10, 11, \cdots, 25]$. More specifically, the experimental setup evaluates performance metrics of the TCR infrastructure for a repository containing $2^{10} = 1024$ to all the way upto $2^{25}$ (32 million) containers stored in the service. A standard mock container image of size 12KB, along with its build and compose code files were used in the simulation. 

For each value of $h$, the database was populated with $2^{h} - N$ records, where $N = 500$, in order to simulate near full load to the repository. The resulting state of the module \textbf{T} was also updated to reflect the state of the repository. In-memory arrays and calculations were used to pre-compute the database records and module state prior to bulk insertion to the database. The pre-populated database was then queried/updated for $N=500$ operations to evaluate the functionality of --- 1) Create ($F_{tp}()$), 2) Update ($F_{tp}()$), 3) Information retrieve ($F_{verify}()$), 4) Encrypted update ($F_{tp}()$ and $F_{st}()$), and 5) Encrypted retrieve ($F_{rs}()$). 

A fine-grained (microsecond resolution) timing scheme was used to record duration for each of the aforementioned operations and their subsequent function calls. Each function simulation was repeated 25 times in order to remove any anomalous behaviour, and median operation times were recorded. The following section (Section \ref{results}) describes the obtained results in greater detail.

\section{Results and Discussion}
\label{results}

\subsection{Performance Evaluation}
\label{perfeval}
\begin{figure}
\centering
\subfloat[Authenticated Create] {
\label{fig:create}
\includegraphics{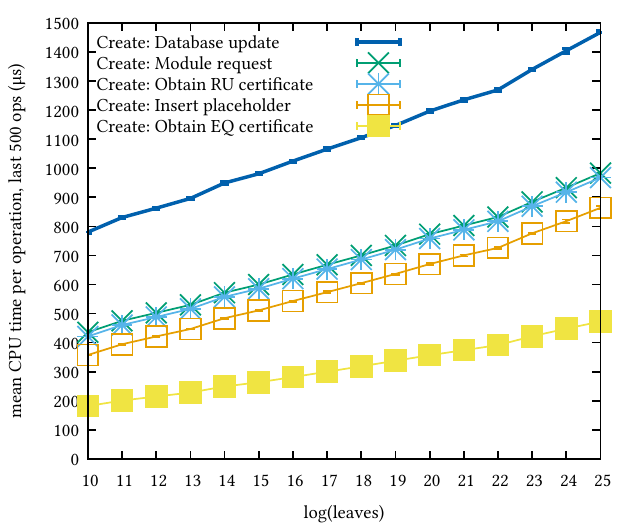}
}
\subfloat[Dummy Create] {
\label{fig:dummycreate}
\includegraphics{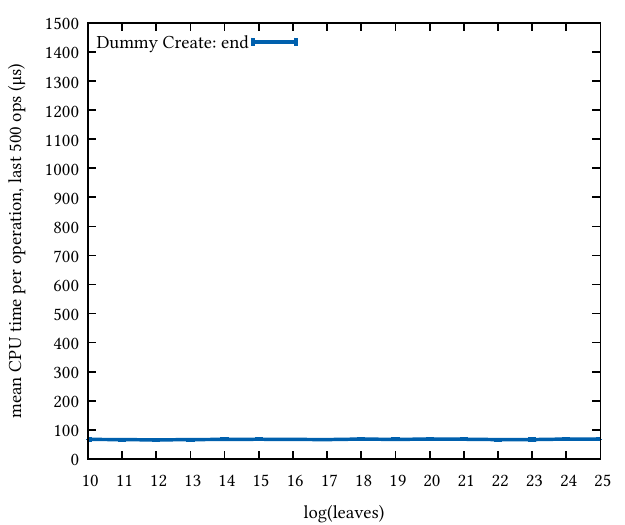}
} \\

\subfloat[Authenticated Modify (unencrypted)] {
\label{fig:modify}
\includegraphics{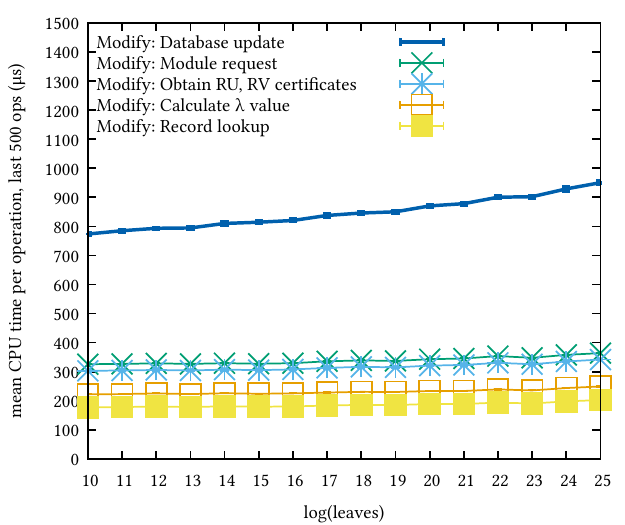}
}
\subfloat[Dummy Modify (unencrypted)] {
\label{fig:dummymodify}
\includegraphics{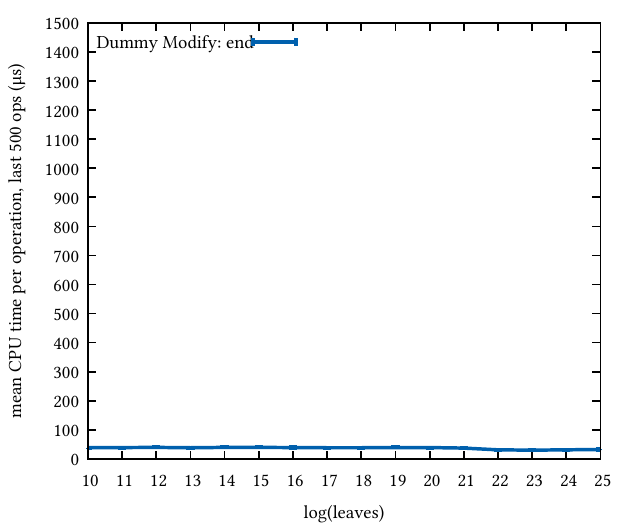}
} \\
\subfloat[Authenticated Retrieve (unencrypted)] {
\label{fig:retrieve}
\includegraphics{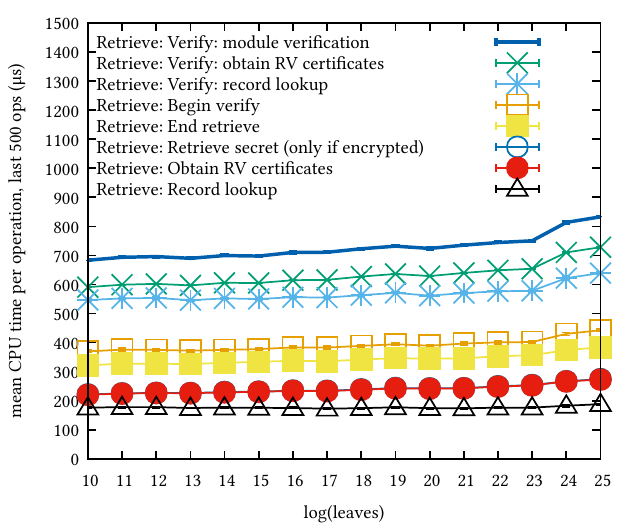}
}
\subfloat[Dummy Retrieve (unencrypted)] {
\label{fig:dummyretrieve}
\includegraphics{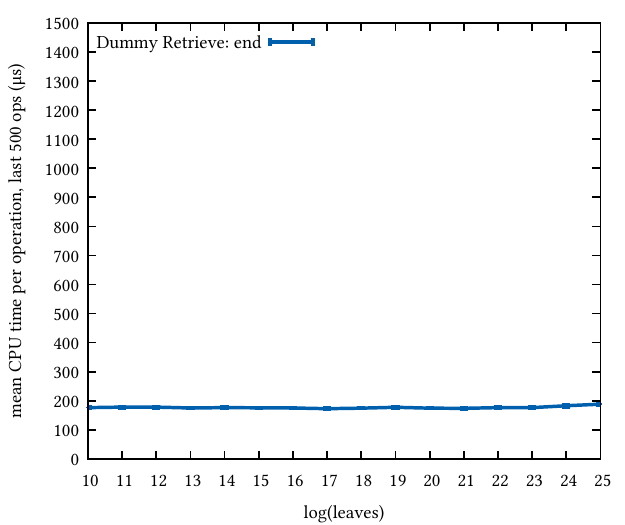}
}
\caption{Average server CPU time (including module time) taken for the last 500 operations by both authenticated and dummy service providers, for various operations. Note the logarithmic x-axis. Error bars show 95\% confidence (+/- 1.96 SE), but may be too small to be visible.}
\end{figure}

Figures \ref{fig:create}-\ref{fig:dummyretrieve} compare the performance of the preliminary TCR implementation described in Section \ref{implementation}, and a dummy unsecure repository. The graphs show average time per operation for the operations of container 1) Creation, 2) Modification, and 3) Retrieval, performed on the repository at varying container loads, given by the tree height $h$ ($n = 2^h$). Across all operations and $h$ values (1024 to 32 million container records), we observe consistent $O(log(n))$ performance, demonstrating the usability and scalability of the TCR architecture.


Container creation (Figure \ref{fig:dummycreate}) takes the greatest amount of time of all tested operations. A worst-case absolute performance of 1.4 ms/operation was observed at a load of $2^{25}$ containers. The steps for container creation (times for each step are displayed in Figure \ref{fig:create}) are 1) generation of an EQ certificate ($F_{eq}()$, Algorithm \ref{f_eq}); 2) insertion of a placeholder for the new container ($F_{ph}()$, Algorithm \ref{f_ph}); 3) obtaining a RU certificate ($F_{ru}()$, Algorithm \ref{f_ru}); 4) root update by \textbf{T} ($F_{tp}()$, Algorithm \ref{f_tp}); and 5) updating service provider database records for the new container.

Compared to a dummy create (Figure \ref{fig:dummycreate}), where the insertion of a record to a unsecure repository takes $O(1)$ time, due to a single database operation of insert), steps 1, 2, and 5 of the create operation take large amounts of time. In step 1 (EQ certificate generation), the model has to look up the complementary nodes ($h+1$ nodes) necessary to perform mapping of an enclosure node to the root node, which requires a database query. Step 2 requires a database update to insert the IOMT leaf record representing the new container. Step 5 requires database updates to insert the new record, increment its counter in the container IOMT, and copy its ACL IOMT into the database. Our preliminary implementation of the database architecture, however, is not optimal (especially for EQ certificate generation), and has potential for optimization.

The time taken for a modification request (Figure \ref{fig:modify}) is broken down into five steps: 1) database lookups for the record; 2) calculating the $\lambda$ value associated with the new container version; 3) obtaining a RU and RV certificate ($F_{ru}()$, Algorithm \ref{f_ru} and $F_{rv}()$, Algorithm \ref{f_rv}, respectively); 4) root update by \textbf{T} ($F_{tp}()$, Algorithm \ref{f_tp}); and 5) update service provider database records. Similar to container creation, the greatest time is taken by step 5, in which database records for the container IOMT, version record table (insertion of a record for modified container), and the container table (container counter) must be updated.

If the user chooses to encrypt the container image, step 5 also includes the time taken by \textbf{T} to decrypt, verify, and re-encrypt a container encryption secret for storage ($F_{st}()$, Algorithm \ref{f_rs}). However, the time taken by this operation is minimal, and is not shown as a separate graph.

In comparison, dummy modify (Figure \ref{fig:dummymodify}) shows constant time for modification (even at different loads), due to the operation only requiring a single database update.

The performance of container retrieval is shown in Figure \ref{fig:retrieve}. In our implementation, a ``retrieval'' is in fact broken down into two separate client-server requests: the first request retrieves the contents of the container image and associated configuration files, and possibly retrieves the encryption key through $F_{rs}()$ (Algorithm \ref{f_rs}); this request, however, does not verify the authenticity of any retrieved data. The second request uses $F_{verify}()$ (Algorithm \ref{f_verify}) to verify the authenticity of data retrieved in the first request.

The times for both requests are shown stacked in one graph in Figure \ref{fig:retrieve}. The steps shown on the graph are 1) database lookups for the requested container and version record; 2) obtaining two RV certificates (one for the container IOMT and one for the ACL IOMT); 3) encryption secret retrieval ($F_{rs}()$, Algorithm \ref{f_rs}); 4) database lookups (again); 5) obtaining two certificates (again); and 6) module \textbf{T} verification ($F_{verify}()$, Algorithm \ref{f_verify}).

Somewhat suboptimally, steps 1--2 and 3--5 are duplicates of each other, resulting from the use of two separate requests for a retrieval query. Performing these steps only once and saving their result would save about 0.2 ms/operation. However, even with this suboptimal implementation, retrieval is still the fastest of all tested operations, with all times being less than 1 ms/operation.

Operations that either modify or retrieve an encrypted container image incur a slight performance penalty over their unencrypted counterparts. Namely, a modification operation with an encrypted container requires \textbf{\textit{S}} to invoke the $F_{st}()$ (Algorithm \ref{f_st}) function exposed by \textbf{T} in order to process the encryption secret for storage, and an encrypted retrieval requires the use of $F_{rs}()$ (Algorithm \ref{f_rs}) to decrypt the encryption secret. However, the time taken by these additional steps is minimal, so separate performance graphs for the two encrypted variants are not shown.

Across all operations, the observed performance demonstrates the exceptional scalability afforded by an IOMT-based design. All operations show nearly logarithmic server performance curves with loads from $2^{10} \approx 10^3$ up to $2^{25} \approx 10^7$ container images. Compared to the 1.5 ms/operation taken by container creation (at $h=25$), modification and retrieval take 0.9 ms/operation and 0.8 ms/operation respectively. Container retrieval requests are, on average, the fastest operation --- this is desirable because a real-world repository usually sees far more retrievals than modifications. This is due to caching of certificates within our databases, leading to constant time query even at higher loads ($h=25$). The certificates invalidate themselves once any updates are performed on the repository and they can no longer be mapped to the root. 

A key limitation of our performance evaluation is that the the service provider \textbf{\textit{S}} and module \textbf{T} are implemented in a single monolithic program. As, there is minimal communication latency between them, while in a real world implementation, \textbf{T} may implemented as an application-specific integrated circuit, which could be orders of magnitude slower than the hardware used by \textbf{\textit{S}}. Thus, the absolute performance figures (the exact time per operation) given by our results may not be representative of real-world numbers. However, the performance \emph{trends} should still hold --- operation times should still scale linearly with $log(n)$, regardless of the absolute performance of \textbf{T} or \textbf{\textit{S}}.

\subsection{TCR Infrastructure Security Assurances}
\label{seceval}

As observed in the previous section, the TCR model is slower in performance comparison to a regular un-secure repository. However, the key benefits of the TCR architecture lies in the security assurances (Section\ref{tcroverview}) provided by the model. Using the authenticated data structure of IOMTs and leveraging the trusted operation of \textbf{T}, \textbf{\textit{S}} the TCR model is able to provide following assurances:  

\begin{itemize}
    \item \textbf{\textit{Integrity}}
    \begin{description}
        \item[\textbf{I1}, \textbf{I2}, \textbf{I3}] - \textit{\textbf{S}} is prevented from successful tampering of any container-related data (image, build code, or deployment code) by TCR model. While the data stored on the servers of \textbf{\textit{S}} can be modified, it would not be able to provide authenticated response of its operations to the users of the service. Successful updates to the container repository (including new container creation), can only be performed by a call to $F_{tp}()$ (Algorithm~\ref{f_tp}) function exposed by \textbf{T}. The certificates generated from $F_{tp}()$ (container and version), cannot be forged by the service provider \textbf{\textit{S}}, as they are signed by the module \textbf{T} secret $\chi$. For a request from user \textbf{\textit{U}} regarding the verification of the container updates, \textbf{\textit{S}} can invoke the function $F_{verify}()$ (Algorithm~\ref{f_verify}), which uses the certificates generated by $F_{tp}()$ in order to provide an authenticated response. The response is generated by \textbf{T} and signed with the module-user secret $K_i$, thwarting any attempts by \textbf{\textit{S}} to manipulate the response. The response from $F_{verify}()$ allows an authorized user to learn the $\lambda$ value associated with any version of a container, which is a commitment to all data related to the container. Replay attacks are also prevented by the inclusion of the request nonce - $\delta$, in the response.
        


        \item[\textbf{I4}] - Only users \textbf{\textit{U}} with sufficient privileges can update/modify container contents. Updates to a container are possible using the $F_{tp}()$, where the access privileges ($a >= 2$ needed for updates) of a user are verified using the ACL IOMT (root $C_{\alpha}$). The function then updates the module root $\xi$ to reflect any changes a privileged user has performed on the repository. Verification of the operation can again be performed using $F_{verify}()$ function. 
        
        
    \end{description}
    \item \textbf{\textit{Availability}}
    \begin{description}
        \item[\textbf{A1}] - \textbf{\textit{S}} cannot deny the existence of any containers if they exist within the repository. The \textbf{\textit{enclosure}}, attributes of the IOMT data structure supports the functionality of authenticated denial. Users can query the status of any container by its index through  $F_{verify}()$. Using the CR certificate of an enclosing leaf (proving non-existence of a leaf index), the module can prove the the users, that the requested container index does not exist, and the response is consistent with the current module root $\xi$. The response to the user contains the queried index and the nonce, signed with $K_i$ to prevent manipulation by \textbf{\textit{S}}.
        
        
        
        \item[\textbf{A2}] - \textbf{\textit{S}} cannot deny the existence of container versions if they exist in TCR. The version record of the containers, store the version counter $C_{VER}$, tracking any updates to the container leading to creation of a new version. Using $F_{verify}()$ function and $CERT_{CR}$ - $CERT_{VR}$ certificates , the module \textbf{T} verifies if there exist any versions of the container ($C_{VER} > 0$). That suggests, if $C_{VER} > 0$, all versions $VER$ such that $1 <= VER <= C_{VER}$ are implied to exist.
        
        
        
    \end{description}
    
    \item \textbf{\textit{Confidentiality}}
    \begin{description}
        \item[\textbf{C1}] - \textbf{\textit{S}} cannot view the contents of the containers as they are encrypted by the key $\sigma$, which is inaccessible to the service provider. The storage of $\sigma$ is handled by the module function call of $F_{st}()$, which encrypts the key using the module secret. The transmission of the secret to the user is done by re-encrypting it with the shared key $K_i$ by \textbf{T}, thereby \textbf{\textit{S}} does not have access to container secret at any point of time.

        \item[\textbf{C2}] - \textbf{\textit{S}} cannot modify the access control list of the containers for malicious intentions. Only an authorized request from an user with access privileges $a >= 3$ can be used to invoke $F_{tp}()$ for updates to ACL. 
        
    \end{description}
        
\end{itemize}

As shown, the attack surface of the TCR model is greatly reduced to the following components --- 1) the trusted hardware boundary of \textbf{T}, 2) operation of TCR functions, and 3) underlying cryptographic functions. The smaller attack surface of the TCR model, enables ease of verifiablity of such components in a complex system. Furthermore, the provable trust in simpler components can then be amplified to assure trust in the entire system. 


\section{Conclusion}
\label{conclusion}

With the current trend towards containerization for most server based software applications, an unsecure centralized container repository opens new attack vectors for potential bad actors due to the implicit trust currently placed in them. Current approaches for securing the distribution of containers, such as Docker Hub, are insufficient due to the potential for improper denial-of-service attacks, which are unacceptable for most mission-critical applications.

The TCR architecture we have presented in our paper aims to address these issues by assuring 1) integrity, 2) availability, and 3) confidentiality, of container images stored in an untrusted service. TCR specifies the requirement of a trusted module \textbf{T}, which provides users with the necessary assurances regarding the repository. Using an authenticated data structure based on index-ordered Merkle trees (IOMTs), and self-certificates/self-memoranda, TCR allows a resource-limited module such as \textbf{T} to efficiently track a virtually unlimited number of containers. 

Additionally, we outline a software implementation of the service provider in a open-source repository \cite{wei} providing proof-of-concept operation of TCR. Performance of such an architecture shows logarithmic server and module time complexity (for container creation, update, and retrieval) at loads from $1024$ up to $33,554,432$ containers. 

While the scalability of the TCR model shows promising results, it leaves quite a bit of room for optimization of its database operations. 
We also plan to explore sub-tree based trust, where trust within internal nodes can be provided without mapping the root everytime. Blockchain technology will also be explored as a transaction based system for keeping track of all operations performed on the repository. Transaction verification incentive in terms of cloud compute/storage credits can be provided for user participation.

\newpage


\bibliographystyle{abbrv}
\bibliography{references.bib}

\begin{thebibliography}{10}

\bibitem{bowers2009hail}
K.~D. Bowers, A.~Juels, and A.~Oprea.
\newblock Hail: A high-availability and integrity layer for cloud storage.
\newblock In {\em Proceedings of the 16th ACM conference on Computer and
  communications security}, pages 187--198. ACM, 2009.

\bibitem{buchmann2008merkle}
J.~Buchmann, E.~Dahmen, and M.~Schneider.
\newblock Merkle tree traversal revisited.
\newblock In {\em International Workshop on Post-Quantum Cryptography}, pages
  63--78. Springer, 2008.

\bibitem{linux_containers}
L.~Containers.
\newblock Infrastructure for container projects.
\newblock \url{http://linuxcontainers.org/}, 2018.

\bibitem{crosby2009efficient}
S.~A. Crosby and D.~S. Wallach.
\newblock Efficient data structures for tamper-evident logging.
\newblock In {\em USENIX Security Symposium}, pages 317--334, 2009.

\bibitem{devanbu2002authentic}
P.~Devanbu, M.~Gertz, C.~Martel, and S.~G. Stubblebine.
\newblock Authentic third-party data publication.
\newblock In {\em Data and Application Security}, pages 101--112. Springer,
  2002.

\bibitem{contenttrust3}
Docker.
\newblock Docker {Content Trust} secures distribution of containerized
  applications.
\newblock
  \url{https://www.docker.com/docker-news-and-press/docker-content-trust-secures-distribution-containerized-applications},
  Aug 2015.

\bibitem{contenttrust}
Docker.
\newblock Content {Trust} in {Docker}.
\newblock \url{https://docs.docker.com/engine/security/trust/content\_trust/},
  2018.

\bibitem{dockerhub}
Docker.
\newblock Docker {Hub}.
\newblock \url{https://hub.docker.com/}, 2018.

\bibitem{dua2014virtualization}
R.~Dua, A.~R. Raja, and D.~Kakadia.
\newblock Virtualization vs containerization to support {PaaS}.
\newblock In {\em Cloud Engineering (IC2E), 2014 IEEE International Conference
  on}, pages 610--614. IEEE, 2014.

\bibitem{erway2015dynamic}
C.~C. Erway, A.~K{\"u}p{\c{c}}{\"u}, C.~Papamanthou, and R.~Tamassia.
\newblock Dynamic provable data possession.
\newblock {\em ACM Transactions on Information and System Security (TISSEC)},
  17(4):15, 2015.

\bibitem{goodin18}
D.~Goodin.
\newblock Backdoored images downloaded 5 million times finally removed from
  {Docker Hub}.
\newblock Ars Technica, Jun 2018.

\bibitem{gummaraju2015over}
J.~Gummaraju, T.~Desikan, and Y.~Turner.
\newblock Over 30\% of official images in {Docker Hub} contain high priority
  security vulnerabilities.
\newblock In {\em Technical Report}. BanyanOps, 2015.

\bibitem{hu2003sead}
Y.-C. Hu, D.~B. Johnson, and A.~Perrig.
\newblock {SEAD}: Secure efficient distance vector routing for mobile wireless
  ad hoc networks.
\newblock {\em Ad hoc networks}, 1(1):175--192, 2003.

\bibitem{kamara2010cryptographic}
S.~Kamara and K.~Lauter.
\newblock Cryptographic cloud storage.
\newblock In {\em International Conference on Financial Cryptography and Data
  Security}, pages 136--149. Springer, 2010.

\bibitem{c_groups}
M.~Kerrisk.
\newblock Linux control groups.
\newblock \url{http://man7.org/linux/man-pages/man7/cgroups.7.html}, 2018.

\bibitem{li2006dynamic}
F.~Li, M.~Hadjieleftheriou, G.~Kollios, and L.~Reyzin.
\newblock Dynamic authenticated index structures for outsourced databases.
\newblock In {\em Proceedings of the 2006 ACM SIGMOD international conference
  on Management of data}, pages 121--132. ACM, 2006.

\bibitem{li2014efficient}
H.~Li, R.~Lu, L.~Zhou, B.~Yang, and X.~Shen.
\newblock An efficient {Merkle-tree}-based authentication scheme for smart
  grid.
\newblock {\em IEEE Systems Journal}, 8(2):655--663, 2014.

\bibitem{li2004secure}
J.~Li, M.~N. Krohn, D.~Mazieres, and D.~E. Shasha.
\newblock Secure untrusted data repository ({SUNDR}).
\newblock In {\em OSDI}, volume~4, pages 9--9, 2004.

\bibitem{martel2004general}
C.~Martel, G.~Nuckolls, P.~Devanbu, M.~Gertz, A.~Kwong, and S.~G. Stubblebine.
\newblock A general model for authenticated data structures.
\newblock {\em Algorithmica}, 39(1):21--41, 2004.

\bibitem{merkel2014docker}
D.~Merkel.
\newblock Docker: lightweight {Linux} containers for consistent development and
  deployment.
\newblock {\em Linux Journal}, 2014(239):2, 2014.

\bibitem{merkle88}
R.~C. Merkle.
\newblock A digital signature based on a conventional encryption function.
\newblock In C.~Pomerance, editor, {\em Advances in Cryptology --- CRYPTO '87},
  pages 369--378, Berlin, Heidelberg, 1988. Springer Berlin Heidelberg.

\bibitem{mohanty14}
S.~Mohanty and M.~Ramkumar.
\newblock Assuring a cloud storage service.
\newblock {\em International Journal of Information Sciences and Computer
  Engineering}, 12 2014.

\bibitem{mohanty16}
S.~Mohanty, M.~Ramkumar, and N.~Adhikari.
\newblock {OMT} : A dynamic authenticated data structure for security kernels.
\newblock {\em International Journal of Computer Networks \& Communications},
  8:1--23, 07 2016.

\bibitem{mohanty11}
S.~D. Mohanty and M.~Ramkumar.
\newblock Securing file storage in an untrusted server-using a minimal trusted
  computing base.
\newblock In {\em CLOSER}, pages 460--470, 2011.

\bibitem{mohanty2012efficient}
S.~D. Mohanty, A.~Velagapalli, and M.~Ramkumar.
\newblock An efficient tcb for a generic content distribution system.
\newblock In {\em Cyber-Enabled Distributed Computing and Knowledge Discovery
  (CyberC), 2012 International Conference on}, pages 5--12. IEEE, 2012.

\bibitem{morris2011trusted}
T.~Morris.
\newblock Trusted platform module.
\newblock In {\em Encyclopedia of cryptography and security}, pages 1332--1335.
  Springer, 2011.

\bibitem{mykletun2006authentication}
E.~Mykletun, M.~Narasimha, and G.~Tsudik.
\newblock Authentication and integrity in outsourced databases.
\newblock {\em ACM Transactions on Storage (TOS)}, 2(2):107--138, 2006.

\bibitem{contenttrust2}
D.~Mónica.
\newblock Introducing {Docker} {Content Trust}.
\newblock \url{https://blog.docker.com/2015/08/content-trust-docker-1-8/},
  2015.

\bibitem{ren2012security}
K.~Ren, C.~Wang, and Q.~Wang.
\newblock Security challenges for the public cloud.
\newblock {\em IEEE Internet Computing}, 16(1):69--73, 2012.

\bibitem{cappos10}
J.~Samuel, N.~Mathewson, J.~Cappos, and R.~Dingledine.
\newblock Survivable key compromise in software update systems.
\newblock In {\em Proceedings of the 17th ACM conference on Computer and
  communications security}, pages 61--72. ACM, 2010.

\bibitem{sarmenta06}
L.~F. Sarmenta, M.~Van~Dijk, C.~W. O'Donnell, J.~Rhodes, and S.~Devadas.
\newblock Virtual monotonic counters and count-limited objects using a tpm
  without a trusted os.
\newblock In {\em Proceedings of the first ACM workshop on Scalable trusted
  computing}, pages 27--42. ACM, 2006.

\bibitem{shu2017study}
R.~Shu, X.~Gu, and W.~Enck.
\newblock A study of security vulnerabilities on {Docker Hub}.
\newblock In {\em Proceedings of the Seventh ACM on Conference on Data and
  Application Security and Privacy}, pages 269--280. ACM, 2017.

\bibitem{tate13}
S.~R. Tate, R.~Vishwanathan, and L.~Everhart.
\newblock Multi-user dynamic proofs of data possession using trusted hardware.
\newblock In {\em Proceedings of the Third ACM Conference on Data and
  Application Security and Privacy}, CODASPY '13, pages 353--364, New York, NY,
  USA, 2013. ACM.

\bibitem{vaughan2006new}
S.~J. Vaughan-Nichols.
\newblock New approach to virtualization is a lightweight.
\newblock {\em Computer}, 11:12--14, 2006.

\bibitem{wan2012hasbe}
Z.~Wan, J.~Liu, and R.~H. Deng.
\newblock Hasbe: A hierarchical attribute-based solution for flexible and
  scalable access control in cloud computing.
\newblock {\em IEEE transactions on information forensics and security},
  7(2):743--754, 2012.

\bibitem{wang2014oruta}
B.~Wang, B.~Li, and H.~Li.
\newblock Oruta: Privacy-preserving public auditing for shared data in the
  cloud.
\newblock {\em IEEE transactions on cloud computing}, 2(1):43--56, 2014.

\bibitem{wang2010privacy}
C.~Wang, Q.~Wang, K.~Ren, and W.~Lou.
\newblock Privacy-preserving public auditing for data storage security in cloud
  computing.
\newblock In {\em Infocom, 2010 proceedings ieee}, pages 1--9. Ieee, 2010.

\bibitem{wang2010hierarchical}
G.~Wang, Q.~Liu, and J.~Wu.
\newblock Hierarchical attribute-based encryption for fine-grained access
  control in cloud storage services.
\newblock In {\em Proceedings of the 17th ACM conference on Computer and
  communications security}, pages 735--737. ACM, 2010.

\bibitem{wei}
F.~Wei.
\newblock Preliminary implementation of {TCR/CSAA}.
\newblock \url{https://github.com/built1n/csaa}, 2018.

\end{thebibliography}
\end{document}